\documentclass[11pt,hyper,a4paper]{article}

\usepackage{jheppub}
\usepackage{amsmath,amssymb,amsfonts,amscd,bm}
\usepackage{graphicx, wrapfig}

\usepackage{soul}
\usepackage{cancel}
\usepackage[normalem]{ulem}
\usepackage{graphicx}
\usepackage{bm}
\usepackage{color}
\usepackage[usenames,dvipsnames,svgnames,table]{xcolor}
\definecolor{darkblue}{cmyk}{0.9,0.9,0,0}
\hypersetup{colorlinks=false, linkcolor=darkblue,linkcolor=darkblue, citecolor=darkblue}

\usepackage{multirow}
\usepackage{verbatim}
\usepackage{url}

\title{Topologically Twisted SUSY Gauge Theory, Gauge-Bethe Correspondence and Quantum Cohomology}
\author{Hee-Joong Chung and Yutaka Yoshida}

\affiliation{Korea Institute for Advanced Study, Seoul}

\abstract{We calculate the partition function and correlation functions in A-twisted 2d $\mathcal{N}=(2,2)$ $U(N)$ gauge theories and topologically twisted 3d $\mathcal{N}=2$ $U(N)$ gauge theories containing an adjoint chiral multiplet with particular choices of $R$-charges and the magnetic fluxes for flavor symmetries. 
According to the Gauge-Bethe correspondence, they correspond to the Heisenberg XXX$_{1/2}$ and XXZ$_{1/2}$ spin chain models, respectively.
We identify the partition function with the inverse of the norm of the Bethe eigenstate. 
Correlation functions are identified to coefficients of the expectation value of Baxter $Q$-operator. 
In addition, we consider correlation functions of 2d $\mathcal{N}=(2,2)^*$ theories and their relations to the equivariant integration of the equivariant quantum cohomology classes of the cotangent bundle of Grassmann manifolds and the equivariant quantum cohomology ring. 
Also, we study the twisted chiral ring relations of supersymmetric Wilson loops in 3d $\mathcal{N}=2^*$ theories and the Bethe subalgebra of the XXZ$_{1/2}$ spin chain models.
\\
\\
\\
\\
\\
\\
\\
{\tt KIAS-P16038}}

\begin{document}

\maketitle


\section{Introduction}
\label{sec:intro}
The Gauge-Bethe correspondence states that quantum integrable models correspond to supersymmetric gauge theories. 
The XXX Heisenberg spin chain model was considered as one of the primary examples of the Gauge-Bethe correspondence in the original papers \cite{NS-GB1,NS-GB2}. 
It was argued that the supersymmetric vacua of the softly broken 2d $\mathcal{N}=(4,4)$ $U(N)$ gauge theory by the mass of the adjoint chiral multiplet, usually called 2d $\mathcal{N}=(2,2)^*$ $U(N)$ gauge theory, is naturally identified with the Bethe ansatz equation for the XXX$_{1/2}$ spin chain model.
Also, the twisted superpotential was identified with the Yang-Yang potential.

Recently, the partition function and correlation functions of topologically twisted 2d $\mathcal{N}=(2,2)$ theories on $S^2$ \cite{CCP-2d} have been calculated.
Also, the partition function of topologically twisted 3d $\mathcal{N}=2$ theories on $S^1 \times S^2$ \cite{Benini-Zaffaroni} (see also \cite{OY-Seifert}) and of topologically twisted 4d $\mathcal{N}=1$ theories on $T^2 \times S^2$ \cite{Closset:2013sxa,Nishioka:2014zpa,Benini-Zaffaroni,Honda:2015yha} have been obtained by considering the rigid limit of supergravity.

In this paper, we study 2d $\mathcal{N}=(2,2)$ and 3d $\mathcal{N}=2$ theories containing an adjoint chiral multiplet with two different choices of $R$-charges and background magnetic fluxes but with same gauge group and matter contents. 
We calculate partition functions of A-twisted 2d $\mathcal{N}=(2,2)$ theories on $S^2$ and partition functions of topologically twisted 3d $\mathcal{N}=2$ theories on $S^1 \times S^2$ with all the equivariant parameters associated to flavor symmetries turned on but with the equivariant parameter associated to the rotational symmetry on $S^2$ turned off. 
We match them with the inverse of the norm of Bethe eigenstates by choosing particular $R$-charges and background fluxes for flavor symmetries. 
The gauge invariant operators form a twisted chiral ring and expectation values of them provide the coefficient of the expectation value of the Baxter $Q$-operator. 
Thus, with a proper choice of coefficients, the expectation value of gauge invariant operators provide the expectation value of conserved charges of the corresponding spin chain model.

We also calculate correlation functions of the A-twisted 2d $\mathcal{N}=(2,2)^*$ theory whose target space (in the nonlinear sigma model limit) is the cotangent bundle of the Grassmannian for several examples.  
We calculate the equivariant integration by using the results in \cite{GRTV} where they showed that the Bethe subalgebra of the XXX spin chain model is isomorphic to the equivariant quantum cohomology ring\footnote{They considered general partial flag manifolds and the Grassmannian is a part of them.}, and check that the result is consistent with correlation functions of the A-twisted 2d $\mathcal{N}=(2,2)^*$ theory and also with the Seiberg-like duality.

It was shown in \cite{RTV} that the Bethe subalgebra of the XXZ spin chain model is given by certain generators and relations analogous to the equivariant quantum cohomology ring in \cite{GRTV}.\footnote{The Bethe subalgebra of the XXZ spin chain model was conjectured to be identical to the equivariant quantum $K$-theory ring \cite{RTV}.} 
With the Gauge-Bethe correspondence in mind, we see that the Wilson loop algebra agrees with the Bethe subalgebra of the XXZ$_{1/2}$ model by checking several examples.
Also, we consider the Seiberg-like duality of the 3d $\mathcal{N}=2^*$ theory in the context of the Bethe subalgebra of the XXZ$_{1/2}$ model.

In the final section, we conclude with a summary of our results and discuss some future directions.


\section{The Gauge-Bethe correspondence and the Bethe norm}
\label{sec:GB}
Given 2d $\mathcal{N}=(2,2)$ gauge theories, the condition for the supersymmetric vacua is given by
\begin{align}
\exp \left( 2\pi i \frac{\partial \widetilde{\mathcal{W}}_{\text{eff}} (\sigma) }{\partial \sigma_a} \right) = 1	\,	, 
\end{align}
where $\widetilde{\mathcal{W}}_{\text{eff}} (\sigma)$ is the effective twisted superpotential.
According to the Gauge-Bethe correspondence \cite{NS-GB1,NS-GB2,NS-GB3}, it is identified with the Bethe ansatz equation of a certain integrable model. 
Also, the twisted superpotential $\widetilde{\mathcal{W}}_{\text{eff}} (\sigma)$ of 2d $\mathcal{N}=(2,2)$ theories corresponds to the Yang-Yang potential of the corresponding integrable model. 
For the isotropic $SU(2)$ Heisenberg XXX$_{1/2}$ spin chain model and similarly for the anisotropic XXZ$_{1/2}$ spin chain model where spin-$1/2$ degree of freedom of $SU(2)$ is attached to each sites, twisted mass parameters for flavor symmetries are related to parameters for the displacement of lattice sites with respect to the symmetric round lattice configuration.

In this section, we relate the norm of the Bethe eigenstates of the XXX$_{1/2}$ and the XXZ$_{1/2}$ spin chain model to the partition function of a certain topologically twisted 2d $\mathcal{N}=(2,2)$ and 3d $\mathcal{N}=2$ theory, respectively. 
We also discuss coefficients of the expectation value of the Baxter $Q$-operator and conserved charges in terms of correlation functions.


\subsection{The norm of the Bethe eigenstate in the $\text{XXX}_{1/2}$ and the $\text{XXZ}_{1/2}$ spin chain model}
\label{ssec:norm}
We are interested in the inhomogeneous $\text{XXX}_{1/2}$ and $\text{XXZ}_{1/2}$ spin chain model with $M$ lattice sites.\footnote{See \cite{QISM, Faddeev:1996iy} for review.} 
The monodromy matrix, $\mathsf{T}(\lambda)$, of the XXX$_{1/2}$ and the $\text{XXZ}_{1/2}$ model takes a form of a $2 \times 2$ matrix
\begin{align}
\mathsf{T}(\lambda) \, = \,  \left( \begin{matrix} \mathsf{A}(\lambda) & \mathsf{B}(\lambda) \\ \mathsf{C}(\lambda) & \mathsf{D}(\lambda) \end{matrix} \right)
\end{align}
acting on the 2-dimensional auxiliary space $V$ where $\lambda$ is a spectral parameter. 
Therefore the transfer matrix, $\tau$, which is given by the trace of monodromy matrix 
\begin{align}
\tau(\lambda) \, = \, \text{Tr} \, \mathsf{T}(\lambda) \, ,
\end{align} 
is $\tau(\lambda) \, = \, \mathsf{A}(\lambda) + \mathsf{D}(\lambda)$. 
With the quasi-periodic boundary condition $\vec{S}_{M+1} = e^{\frac{i}{2} \vartheta \sigma_3} \vec{S}_1 e^{-\frac{i}{2} \vartheta \sigma_3}$ where $\vec{S}_a = \frac{1}{2}\vec{\sigma}_{(a)}$ are generators at the $a$-th site and $\vec{\sigma}$ are the Pauli matrices, the transfer matrix is given by $\mathsf{A}(\lambda)+ e^{i\vartheta} \mathsf{D}(\lambda)$ \cite{DEVEGA1984495, NS-GB1}.

The $\mathsf{R}$-matrix of the XXX$_{1/2}$ and the XXZ$_{1/2}$ model is 
\begin{align}
\mathsf{R}(\lambda, \mu) \, = \, \left(  \begin{matrix}
f(\mu, \lambda)	& 0 				& 0				& 0	\\
0			& g(\mu, \lambda)	& 1				& 0	\\
0			& 1				& g(\mu, \lambda)	& 0	\\
0			& 0				& 0				& f(\mu, \lambda) 
\end{matrix}
\right)
\end{align}
with
\begin{align}
f(\mu,\lambda) \, = \, 1 + \frac{ic}{\mu-\lambda}	\,	,	\quad g(\mu,\lambda) \, = \, \frac{ic}{\mu-\lambda} \, 
\end{align}
for the XXX$_{1/2}$ model where $c$ is an auxiliary parameter, and
\begin{align}
f(\mu, \lambda) = \frac{\sinh (\mu - \lambda + 2i\eta)}{\sinh(\mu - \lambda)}	\,	,	\quad g(\mu, \lambda) = \frac{i \sin 2 \eta}{ \sinh(\mu - \lambda)}
\end{align}
for the XXZ$_{1/2}$ model where $\eta$ is related to the anisotropy parameter $\Delta = \cos 2 \eta$, $0 < 2 \eta \leq \pi$.

The $\mathsf{R}$-matrix satisfies the Yang-Baxter equation 
\begin{align}
\mathsf{R}_{12}(\lambda, \mu) \mathsf{R}_{13}(\lambda, \nu) \mathsf{R}_{23}(\mu, \nu) \, = \, \mathsf{R}_{23}(\mu, \nu) \mathsf{R}_{13}(\lambda, \nu) \mathsf{R}_{12}(\lambda, \mu)
\end{align}
acting on the auxiliary space $V_1 \otimes V_2 \otimes V_3$ where $\mathsf{R}$-matrix $\mathsf{R}_{ab}$ acts on $V_a \otimes V_b$.
The Yang-Baxter equation implies
\begin{align}
\mathsf{R}_{ab}(\lambda, \mu) (\mathsf{T}_a(\lambda) \otimes \mathsf{T}_b(\mu)) \, = \, (\mathsf{T}_b(\mu) \otimes \mathsf{T}_a(\lambda)) \mathsf{R}_{ab}(\lambda, \mu) \, , \label{YBE2}
\end{align}
where $\mathsf{T}_a$ acts on the auxiliary space $V_a$, and this provides the commutation relations of matrix elements of the monodromy matrix $\mathsf{T}(\lambda)$. 
Also, from \eqref{YBE2} and due to the trace identities, one can show that the transfer matrix, $\tau(\lambda)$, commutes with the Hamiltonian, 
\begin{align}
[ \tau(\lambda) \, , \, H] \, = \, 0	\,	.
\end{align}
Therefore $\tau(\lambda)$ is a generating function of conserved charges. 
As they commute, eigenfunctions of the transfer matrix are also eigenfunctions of the Hamiltonian.

The pseudo-vacuum $|0\rangle$ satisfies the following conditions
\begin{align}
\mathsf{A}(\lambda) |0\rangle \, = \, \mathsf{a}(\lambda)|0\rangle	\,	,	\quad	
\mathsf{D}(\lambda)|0\rangle \, = \, \mathsf{d}(\lambda) |0\rangle	\,	,	\quad
\mathsf{C}(\lambda) |0\rangle \, = \, 0	\,	
\end{align}
where $\mathsf{a}(\lambda)$ and $\mathsf{d}(\lambda)$ are called the vacuum eigenvalues. 
For the Heisenberg spin chain model, the pseudo-vacuum $|0\rangle$ is given by the state with spins being all up or all down.


\subsubsection*{The Bethe eigenstate}
Consider a state that is obtained by acting an operator $B$ on the pseudo-vacuum
\begin{align}
| \Psi_N ( \lambda)\rangle \, = \,  \prod_{a=1}^N \, B(\lambda_a) |0 \rangle
\end{align}
where $N$ is the number of particles or excitations. 
This state becomes the eigenvector of the transfer matrix when spectral parameters, $\lambda_a$, satisfy the Bethe ansatz equation, and the eigenvector is called the Bethe eigenstate. 
The dual vector of $| \Psi_N ( \lambda)\rangle$ is defined by
\begin{align}
\langle  \Psi_N (\lambda )| \, = \,  \langle 0 | \prod_{a=1}^N \, C(\lambda_a)	\,	. 
\end{align}

The vacuum eigenvalues, $\mathsf{a}(\lambda)$ and $\mathsf{d}(\lambda)$, of the inhomogeneous XXX$_{1/2}$ and XXZ$_{1/2}$ model are
\begin{align}
\mathsf{a}(\lambda) = \prod_{j=1}^{M} \left(\lambda - \nu_j -i\frac{c}{2} \right), \quad 
\mathsf{d}(\lambda) = \prod_{j=1}^{M} \left(\lambda - \nu_j + i\frac{c}{2} \right)	\,	,
\end{align}
and
\begin{align}
\mathsf{a}(\lambda) = \prod_{j=1}^M \sinh(\lambda - \nu_j - i \eta), \quad \mathsf{d}(\lambda) = \prod_{j=1}^M \sinh(\lambda - \nu_j + i \eta) \, ,  \label{vacEvalXXZ}
\end{align}
respectively, and the Bethe ansatz equation is
\begin{align}
\prod_{j=1}^M \frac{\lambda_a - \nu_j -i\frac{c}{2}}{\lambda_a - \nu_j + i\frac{c}{2}} 
\, = \, e^{i\vartheta} \, \prod_{b =1 \atop b \neq a}^{N} \frac{\lambda_b- \lambda_a + ic }{\lambda_b- \lambda_a - ic } \, ,
\label{BeqXXX}
\end{align}
and
\begin{align} 
\prod_{j=1}^M \frac{\sinh(\lambda_a - \nu_j - i \eta)}{\sinh(\lambda_a - \nu_j + i \eta)} \,  =  \, e^{i\vartheta} \, \prod_{b =1 \atop b\neq a}^N \frac{\sinh(\lambda_b-\lambda_a+2i\eta)}{\sinh(\lambda_b-\lambda_a-2i\eta)} \, , 
\label{BeqXXZ}
\end{align}
respectively, for the quasi-periodic boundary condition.


\subsubsection*{The norm of the Bethe eigenstate for XXX$_{1/2}$ model}
The norm of the Bethe eigenstate \cite{korepin1982} is given by
\begin{align}
\langle \Psi_N( \lambda )| \Psi_N( \lambda )\rangle = c^N \prod_{a=1}^{N} \mathsf{a}(\lambda_a) \, \mathsf{d}(\lambda_a)  \, \prod_{a < b} f(\lambda_a, \lambda_b) f(\lambda_b, \lambda_a) \; \det(\varphi^{'})
\end{align}
where
\begin{align}
\varphi^{'}_{ab} &= \delta_{ab} \Big(  i \frac{\partial}{\partial \lambda_a} \log r(\lambda_a) + \sum_{l=1}^N K(\lambda_a, \lambda_l) \Big) - K(\lambda_a, \lambda_b) \, , \\
&K(\lambda, \mu) = \frac{2c}{(\lambda - \mu)^2 + c^2}, \quad r(\lambda) = \frac{\mathsf{a}(\lambda)}{\mathsf{d}(\lambda)}	\,	.
\end{align}
For the inhomogeneous XXX$_{\frac{1}{2}}$ spin chain model, 
\begin{align}
\begin{split}
\varphi_{ab}^{'} =& \; i \delta_{ab}  \Bigg[ \Bigg(  \sum_{l=1}^M \bigg( \frac{1}{\lambda_a - \nu_l -i \frac{c}{2}} - \frac{1}{\lambda_a - \nu_l + i \frac{c}{2}} \bigg) + \sum_{s=1}^N \bigg( \frac{1}{\lambda_s-\lambda_a+ic} +  \frac{1}{\lambda_a-\lambda_s+ic} \bigg) \Bigg) \\
& \qquad \quad  - \bigg(  \frac{1}{\lambda_a-\lambda_b+ic} + \frac{1}{\lambda_b-\lambda_a+ic} \bigg) \Bigg] =: i \widetilde{\varphi}^{'}_{ab} \, ,
\end{split}
\end{align}
and we obtain
\begin{align}
\hspace{-5mm} \langle \Psi_N(\lambda )| \Psi_N( \lambda )\rangle = c^N 
\prod_{a=1}^{N} \prod_{j=1}^M \left(\lambda_a - \nu_j -i \frac{c}{2} \right) \left(\lambda_a - \nu_j + i \frac{c}{2} \right)
\prod_{a \neq b} \frac{(\lambda_a-\lambda_b+ic)}{(\lambda_b-\lambda_a)} \; \det(\varphi^{'}) \, .
\end{align}
Therefore, the inverse of the norm of the Bethe eigenstate is given by
\begin{align}
\begin{split}
&\sum_{ ( \lambda ) \in P_{\text{XXX}}} \langle \Psi_N( \lambda )| \Psi_N( \lambda )\rangle^{-1} \\
 & ~~~~~~~~=\sum_{ ( \lambda )  \in P_\text{XXX}}  (ic)^{-N} \prod_{a \neq b} \frac{\lambda_a - \lambda_b}{\lambda_a-\lambda_b+ic} 
\prod_{a=1}^{N} \prod_{j=1}^M \frac{1}{(\lambda_a - \nu_j -i \frac{c}{2}) (\lambda_a - \nu_j + i \frac{c}{2})}\;  \det(\widetilde{\varphi}^{'})^{-1}	\,	.
\end{split}
\label{eq:inverseXXX}
\end{align}
Here $P_{\text{XXX}}$ is a set of independent solutions of $(\lambda) := (\lambda_{1}, \cdots, \lambda_{N})$ satisfying the Bethe ansatz \eqref{BeqXXX} with the quasi-periodic boundary condition.


\subsubsection*{The norm of the Bethe eigenstate for XXZ$_{1/2}$ model}
The norm of the Bethe eigenstate for the XXZ$_{1/2}$ model \cite{korepin1982} can be obtained similarly as in the case of the XXX$_{1/2}$ model and it is 
\begin{align}
\begin{split}
\langle \Psi_N( \lambda )| \Psi_N(\lambda )\rangle =& (\sin 2\eta)^N 
 \prod_{a \neq b} \frac{\sinh(\lambda_a - \lambda_b + 2i\eta)}{\sinh(\lambda_a - \lambda_b)} \\
&\hspace{17mm} \times  \prod_{a=1}^N \prod_{j=1}^M \sinh (\lambda_a - \nu_j -i\eta) \sinh (\lambda_a - \nu_j + i\eta) \det (\varphi^\prime)
\end{split}
\end{align}
where
\begin{align}
\begin{split}
&\hspace{-17mm} \varphi_{ab}^\prime  = i \delta_{ab} \left[   \sum_{j=1}^M \left( \frac{\cosh(\lambda_a - \nu_j - i \eta)}{\sinh(\lambda_a - \nu_j - i \eta)} - \frac{\cosh(\lambda_a - \nu_j + i \eta)}{\sinh(\lambda_a - \nu_j + i \eta)} \right) +  \sum_{e=1}^N \left( \frac{\cosh(\lambda_a - \lambda_e + 2i \eta)}{\sinh(\lambda_a - \lambda_e + 2i \eta)} - \frac{\cosh(\lambda_a - \lambda_e - 2i \eta)}{\sinh(\lambda_a - \lambda_e - 2i \eta)} \right) \right] \\
&   -i\left( \frac{\cosh(\lambda_a - \lambda_b + 2i \eta)}{\sinh(\lambda_a - \lambda_b + 2i \eta)} - \frac{\cosh(\lambda_a - \lambda_b - 2i \eta)}{\sinh(\lambda_a - \lambda_b - 2i \eta)} \right)\\
&=: i \widetilde{\varphi}^{'}_{ab}	\,	.
\end{split}
\end{align}
In terms of $\widetilde{\varphi}'$, the inverse of the norm of the Bethe eigenstate is given by
\begin{align}
\begin{split}
& \sum_{ (\lambda)  \in P_{\text{XXZ}} }  \langle \Psi_N( \lambda )| \Psi_N( \lambda )\rangle^{-1} =  
\sum_{ (\lambda) \in P_\text{XXZ}} (i\sin 2\eta)^{-N} 
\prod_{a \neq b} \frac{ \sinh(\lambda_a - \lambda_b)}{\sinh(\lambda_a - \lambda_b + 2i\eta) }  
\\
&\hspace{60mm} \times \prod_{a=1}^N \prod_{j=1}^M \frac{1}{\sinh (\lambda_a - \nu_j -i\eta) \sinh (\lambda_a - \nu_j + i\eta)} 
    \big( \det {\widetilde{\varphi}}' \big)^{-1} \,  , 
\end{split}
\end{align}
and $P_{\text{XXZ}}$ is a set of independent solutions of $(\lambda) := (\lambda_{1}, \cdots, \lambda_{N})$ satisfying the Bethe ansatz equation \eqref{BeqXXZ} with the quasi-periodic boundary condition.


\subsection{Correlation functions in the 2d $\mathcal{N}=(2,2)$ gauge theory and the XXX$_{1/2}$ spin chain model}
\label{ssec:GLSM1}
We consider topologically twisted $\mathcal{N}=(2,2)$ $U(N_c)$ gauge theories.
The matter chiral multiplets contain an adjoint chiral multiplet $\Phi$, $N_f$ fundamental chiral multiplets $Q^{a}_{\ i}$ and $N_f$ anti-fundamental chiral multiplets $\widetilde{Q}_{a}^{\ i}$, $a=1, \ldots, N_c$, $i=1, \ldots, N_f$.
The flavor symmetry group is $SU(N_f)_{Q} \times  SU(N_f)_{\widetilde{Q}} \times U(1)_D$.   
The charge assignment is specified in Table \ref{tabel:glsm1}.

\begin{table}[h]
\begin{center}
\begin{tabular}{c | c c c  c c}
			&	$U(N_c)$			&	$SU(N_f)_{{Q}}$		&	$SU(N_f)_{\widetilde{Q}}$	&	$U(1)_D$	&	$U(1)_R$	\\
			\hline
$Q$			&	$N_c$			&	$\overline{N_f}$			&	$\mathbf{1}$					&	$-1/2$			&	$r_1$		\\
$\widetilde{Q}$	&	$\overline{N}_c$	&	$\mathbf{1}$				&	$N_f$				&	$-1/2$	&	$r_2$		\\
$\Phi$		&	\text{adj}			&	 $\mathbf{1}$				&	$\mathbf{1}$					&	$1$				&	$R$
\end{tabular} 
\caption{The matter contents and charge assignment}
\label{tabel:glsm1}
\end{center}
\end{table}
The mass parameters and fluxes of the Cartan of global symmetries are denoted by\footnote{In our notation, the background flux $l$ is an even number.}
\begin{align}
SU(N_f)_Q \, : \, (m^y_i, n_i)	\,	,	\quad 
SU(N_f)_{\widetilde{Q}},   \, : \, (m^{\widetilde{y}}_i, \widetilde{n}_i)	\,	,	\quad 
U(1)_D \, : \, (m^z, l)	\,	.
\end{align}

The partition function of the A-type topologically twisted theory can be calculated by using the formula in \cite{CCP-2d}. 
In the following calculation, we turn off the background value of the graviphoton associated to $S^2$.

The one-loop contributions from the chiral, anti-chiral, and adjoint chiral multiplets are given by
\begin{align}
Z_{Q}^{\text{1-loop}}(k) \, &= \, \prod_{a=1}^{N_c} \prod_{i=1}^{N_f}(\sigma_a - m_i^y - \frac{1}{2}m^z)^{r_1-k_a-1+n_i+\frac{1}{2}l}	\,	,	\\
Z_{\widetilde{Q}}^{\text{1-loop}}(k) \, &= \, \prod_{a=1}^{N_c} \prod_{i=1}^{N_f} (-\sigma_a + m_i^{\widetilde{y}}- \frac{1}{2}m^z)^{r_2+k_a-1-\widetilde{n}_i+\frac{1}{2}l}	\,	,	\\
Z_{\Phi}^{\text{1-loop}}(k) \, &= \, (m^z)^{N(R-1-l)}  \prod_{1 \le a\neq b \le N_c} (\sigma_a - \sigma_b + m^z)^{R-(k_a-k_b)-1- l} 
\end{align}
and the one-loop contribution from the vector multiplet is
\begin{align}
Z^{\text{1-loop}}_{\text{vector}}(k) \, = \,  (-1)^{\frac{N_{c}(N_c-1)}{2}}\prod_{1 \leq a < b \leq N_c} (-1)^{k_a-k_b+1} (\sigma_a - \sigma_b)^2	\,	.
\end{align}
We denote a constant configuration of the scalar in the vector multiplet as $\sigma=\mathrm{diag}(\sigma_1, \cdots, \sigma_{N_c})$. 
The partition function of A-twisted gauged linear sigma models on $S^2$ is given by
\begin{align}
Z^{2d} \, = \, \frac{1}{N_c!} \sum_{\vec{k} \in \mathbb{Z}^{N_c} } q^{\sum_{a=1}^{N_c} k_a} \, \oint \prod_{a=1}^{N_c} \frac{d \sigma_a}{2\pi i}   
\, Z^{\text{1-loop}}_{\text{total}} (k) \, .
\end{align}
Here the choice of the contour is specified by the Jeffrey-Kirwan residue prescription, which depends on the choice of the covector $\eta \in \mathbb{R}^{N_c}$. The parameter $q$ is the exponential of the complexified FI-parameter $q \, := \, \exp {2\pi i \tau} \, = \, \exp {2\pi i \left( \frac{\theta}{2\pi} + i \xi \right)}$ and $Z^{\text{1-loop}}_{\text{total}} (k)$ is
\begin{align}
Z^{\text{1-loop}}_{\text{total}} (k) \, = \, Z^{\text{1-loop}}_{\text{vector}}(k) \, Z_{Q}^{\text{1-loop}}(k) \, Z_{\widetilde{Q}}^{\text{1-loop}}(k) \, Z_{\Phi}^{\text{1-loop}}(k) \, .
\end{align}
By choosing a covector $\eta$, for example, to be $\eta=(-1, \cdots, -1)$, the one-loop determinants of anti-chiral multiplets and an adjoint chiral multiplet contribute to Jeffrey-Kirwan residues.
Poles from anti-chiral multiplets exist when $k_a <  \widetilde{n}_i - \frac{1}{2}l -r_2 +1$. 
Summing over $k_a < K$ first for a sufficiently large positive integer $K$, the partition function is expressed as
\begin{align}
\begin{split}
\hspace{-10mm} Z^{2d} \, = \, \frac{ (-1)^{\frac{N_c(N_c-1)}{2}}   }{N_c !}  \oint \left( \prod_{a=1}^{N_c} \frac{d \sigma_a}{2\pi i} \right) & \,  \prod_{a=1}^{N_c} \frac{\exp(2\pi i \partial_{\sigma_a} \, \widetilde{\mathcal{W}}_{\text{eff}})^K}{\exp(2\pi i \partial_{\sigma_a} \, \widetilde{\mathcal{W}}_{\text{eff}})-1} \, \prod_{1 \leq a<b \leq N_c} (\sigma_a - \sigma_b)^2 \, \prod_{1 \leq a \neq b \leq N_c} (\sigma_a - \sigma_b + m^z)^{R-l-1}  \\
\times \, & (m^z)^{N_c(R-l-1)}  \, \prod_{a=1}^{N_c} \prod_{i=1}^{N_f} (\sigma_a - m^y_i - \frac{1}{2} m^z)^{r_1+n_i+\frac{l}{2}-1} (-\sigma_a + m^{\widetilde{y}}_i - \frac{1}{2} m^z)^{r_2 - \widetilde{n}_i+\frac{l}{2}-1}
\end{split}
\end{align}
where $\widetilde{\mathcal{W}}_{\text{eff}}$ is the effective twisted superpotential
\begin{align}
\begin{split}
&\hspace{-10mm} \widetilde{\mathcal{W}}_\text{eff}  \; = \; \tau \sum_{a=1}^{N_c} \sigma_a -\frac{1}{2} \sum_{1 \leq a < b \leq N_c} (\sigma_a - \sigma_b) \\
&\hspace{-10mm} -\frac{1}{2\pi i }  \Bigg[  \sum_{a=1}^{N_c} \sum_{i=1}^{N_f} (\sigma_a - m_i^y -\frac{1}{2}m^z) (\log (\sigma_a - m_i^y -\frac{1}{2}m^z) -1) + (-\sigma_a + m_i^{\widetilde{y}} -\frac{1}{2}m^z) (\log (-\sigma_a + m_i^{\widetilde{y}} -\frac{1}{2}m^z) -1) \\
& \hspace{90mm}+ \sum_{a,b=1}^{N_c} (\sigma_a - \sigma_b + m^z) (\log (\sigma_a - \sigma_b +m^z) -1)	\Bigg]
\end{split} \label{twst-s-pot}
\end{align}
and
\begin{align}
\exp(2\pi i \partial_{\sigma_a} \, \widetilde{\mathcal{W}}_{\text{eff}}) \, = \, (-1)^{N_c-1}  q \prod_{i=1}^{N_f} \frac{-\sigma_a + m^{\widetilde{y}_i}  -\frac{1}{2}m^z }{\sigma_a - m^y_i - \frac{1}{2}m^z} \, \prod_{ b \neq a} \frac{\sigma_b-\sigma_a+m^z}{\sigma_a-\sigma_b+m^z}	\, .
\end{align}
Due to the factor $\exp(2\pi i \partial_{\sigma_a} \, \widetilde{\mathcal{W}}_{\text{eff}})^K$ with large $K$ in numerator, there are no poles at $-\sigma_a + m^{\widetilde{y}}_i - \frac{1}{2} m^z=0$ and $\sigma_a - \sigma_b + m^z=0$ and only poles at $\exp(2\pi i \partial_{\sigma_a} \, \widetilde{\mathcal{W}}_{\text{eff}})-1=0$ contribute. Then, dependence on $K$ disappears and we obtain
\begin{align}
\begin{split}
Z^{2d} =&(-1)^{\frac{N_c(N_c+1)}{2}}    { (m^z)^{N_c(R-1-l)}} \sum_{( \sigma ) \in P_{2d}}  \det (\mathcal{M}^{2d})^{-1}   
\prod_{a\neq b} (\sigma_a - \sigma_b) (\sigma_a - \sigma_b + m^z)^{R-1- l} \\
&\hspace{30mm} \times \prod_{a=1}^{N_c} \prod_{i=1}^{N_f}(\sigma_a - m_i^y - \frac{1}{2}m^z)^{r_1-1+n_i+\frac{1}{2}l}\,  (-\sigma_a + m_i^{\widetilde{y}}- \frac{1}{2}m^z)^{r_2-1-\widetilde{n}_i+\frac{1}{2}l}	 
\label{2dpartftn}
\end{split}
\end{align}
where
\begin{align}
&P_{\text{2d}} \, := \,  \{ (\sigma_1,\cdots, \sigma_{N_c})  \,  | \, \exp(2 \pi i \partial_{\sigma_a} \widetilde{\mathcal{W}}_\text{eff}) = 1 \; \text{for all}\; a=1, \ldots, N_c \} /S_{N_c} \\
&
\mathcal{M}^{2d}_{ab} \, :=  \, (-2\pi i) \partial_{{\sigma}_a} \partial_{{\sigma}_b} \widetilde{\mathcal{W}}_{\text{eff}} 
\end{align}
and
\begin{align}
\hspace{-4mm} 
-2\pi i \, \partial_{\sigma_a} \partial_{\sigma_b} \widetilde{\mathcal{W}}_{\text{eff}} \, = \, \delta_{ab} \bigg( \frac{1}{\sigma_a - m^y_i -\frac{1}{2}m^z} + \frac{1}{-\sigma_a + m^{\widetilde{y}}_i -\frac{1}{2}m^z} + \sum_{l=1}^{N_c} (\mathcal{S}_{lb} - \mathcal{S}_{bl}) \bigg) - \mathcal{S}_{ab} - \mathcal{S}_{ba} 
\end{align}
with 
\begin{align}
\mathcal{S}_{kl} = \frac{1}{\sigma_{k} - \sigma_l + m^z}	\,	 .
\end{align}
In $P_{\text{2d}}$, we identify solutions which are the same up to Weyl permutations, $S_{N_c}$, of $(\sigma_1, \cdots, \sigma_{N_c})$. 
And the condition for supersymmetric vacua, 
$\exp (2\pi i \partial_{\sigma_a} \widetilde{\mathcal{W}}_{\text{eff}})=1$, is given by
\begin{align}
\prod_{i=1}^{N_f} \frac{(\sigma_a - m_i^y -\frac{1}{2}m^z)}{(\sigma_a - m_i^{\widetilde{y}} +\frac{1}{2}m^z)} = (-1)^{N_f} e^{2\pi i \tau} \prod_{b\neq a}^{N_c} \frac{\sigma_b-\sigma_a +m^z}{\sigma_b-\sigma_a  - m^z} \, . \label{preBetheeq}
\end{align}


\subsection*{Comparison and matching}
The following identifications  
\begin{align}
\sigma / m  = \lambda	\,	,	\quad
m^z / m = ic	\,	, 	\quad  
m^y / m = m^{\widetilde{y}} / m = \nu	\,	,	\quad 
N_c=N	\,	,	\quad 
N_f=M	\,	,	\quad 
(-1)^{N_f}q= e^{i \vartheta}
\end{align}
give the agreement between Bethe ansatz equation \eqref{BeqXXX} for the XXX$_{1/2}$ spin chain model and the condition for supersymmetric vacua \eqref{preBetheeq} of the 2d $\mathcal{N}=(2,2)$ gauge theory where $m$ is an arbitrary parameter with mass dimension one. 
Moreover, with identifications
\begin{align}
r_1 + n_i +\frac{l}{2} = 0	\,	,	\quad 
r_2 - \widetilde{n}_i + \frac{l}{2} = 0	\,	,	\quad 
R-l=0	\,	, 
\label{condXXX}
\end{align}
the partition function of the A-twisted 2d $\mathcal{N}=(2,2)$ gauge theory\footnote{For example, we can choose all background fluxes and $R$-charges to be zero and don't include the superpotential $\widetilde{Q} \Phi Q$ in the theory. 
The canonical assignment of the $R$-charge for superpotential $\widetilde{Q} \Phi Q$ is not allowed if we want to match the A-twisted partition function and the inverse of the norm of the Bethe eigenstate. 
Indeed if we sum three conditions in \eqref{condXXX}, we obtain
\begin{align}
N_f (r_1+r_2+R)+\sum_{i=1}^{N_f}n_i - \sum_{i=1}^{N_f}\widetilde{n}_i \, = \, 0	\,	.
\end{align}
However, as flavor symmetries are $SU(N_f)$ instead of $U(N_f)$, we have $r_1+r_2+R=0$. 
Therefore, the canonical assignment of the $R$-charge such as $r_1=r_2=0$ and $R=2$ is not allowed for the match with the inverse of the norm. 
Also note that, given same matter contents, the Bethe ansatz equation is same whatever the $R$-charges and background magnetic fluxes are.} 
and the inverse of the norm of the Bethe eigenstate \eqref{eq:inverseXXX} agree
\begin{align}
Z^{2d} = \sum_{ ( \lambda ) \in P_{\text{XXX}}} \langle \Psi_N(\lambda )| \Psi_N( \lambda )\rangle^{-1} 
\end{align}
up to an overall factor.
This type of relation was first studied for the $U(N)/U(N)$ gauged WZW model on genus-$g$ Riemann surfaces $ \Sigma_g$ in \cite{Okuda:2012nx} where the corresponding integrable model is the phase model. 
See also \cite{Okuda-Yoshida, Gukov:2015sna, Okuda:2015yea, NS-curved}.


\subsection*{Correlation functions, the Baxter $Q$-operator, and conserved charges}
We have identified the partition function and the norm of the inverse of the Bethe eigenstate. 
We can also consider correlation functions of the A-twisted 2d $\mathcal{N}=(2,2)$ theory discussed in section \ref{ssec:GLSM1} in the context of the Gauge-Bethe correspondence.

In the A-twisted 2d $\mathcal{N}=(2,2)$ theory, correlation functions of gauge invariant operators $\mathcal{O}(\sigma)$ are given by
\begin{align}
\langle \mathcal{O}(\sigma) \rangle \, = \, \frac{1}{N_c!} \sum_{\vec{k} \in \mathbb{Z}^{N_c}} (-1)^{(N_c-1) \sum_{a=1}^{N_c} k_a} q^{\sum_{a=1}^{N_c} k_a} 
\, \oint \prod_{a=1}^{N_c} \frac{d \sigma_a}{2\pi i} \, \mathcal{O}(\sigma)
\, Z^{\text{1-loop}}_{\text{total}} (k) \, .
\end{align}
This can also be written as 
\begin{align}
\langle \mathcal{O}(\sigma) \rangle \, = \sum_{( \sigma ) \in P_{\text{2d}} }  \mathcal{O}(\sigma) \, \frac{Z^{\text{1-loop}}_{\text{total}} (k=0)}
{\mathrm{det} \, \mathcal{M}^{2d}  }	\,	.
\end{align}
The operator $\mathcal{O}(\sigma)$ is provided by gauge invariant polynomials of the Cartan of the scalar component $\sigma$ of the vector multiplet, which is a symmetric function of $\sigma_a$, $a=1, \cdots, N_c$. 
Thus it can be written in terms of the elementary symmetric polynomials. 
We denote the polynomial $Q(x)$ as
\begin{align}
Q(x) \, = \, \prod_{a=1}^{N_c} (x-\sigma_a) \, ,  \label{BaxterQ}
\end{align}
then the coefficients of $x^{N_{c}-l}$ provide the $l$-th elementary symmetric polynomial of $\sigma_a$.

Meanwhile, in integrable models there is a fundamental quantity known as the Baxter $Q$-operator $\mathbf{Q}(x)$ whose eigenvalue is actually \eqref{BaxterQ} with $N_c$ identified with the number of particles $N$ and $\sigma_a$ with spectral parameters $\lambda_a$. 
Thus, we see that the expectation value of the Baxter $Q$-operator provides the generating function of correlation functions of gauge invariant operators in the 2d $\mathcal{N}=(2,2)$ theory in section \ref{ssec:GLSM1}, \textit{i.e.}
\begin{align}
\begin{split}
  \sum_{ ( \lambda ) \in P_\text{XXX}} \frac{ \langle  \Psi_N ( \lambda )| \mathbf{Q}(x) | \Psi_N( \lambda )\rangle}
{ \langle \Psi_N( \lambda )| \Psi_N( \lambda )\rangle^2}  
& = \sum_{(\lambda) \in P_{\text{XXX}}}  (ic)^{-N}  \, Q(x)  \det{^{-1}}(\widetilde{\varphi}^{'})\,  \\
&\hspace{15mm} \times \prod_{a \neq b} \frac{ \lambda_a - \lambda_b}{\lambda_a-\lambda_b+ic }  
\prod_{a=1}^{N} \prod_{j=1}^M \frac{1}{(\lambda_a - \nu_j -i \frac{c}{2}) (\lambda_a - \nu_j + i \frac{c}{2})}\,   .
\end{split}
\end{align}
The eigenvalue of the transfer matrix $\tau(\mu)$ for the XXX$_{1/2}$ model is given by
\begin{align}
\theta \left(\mu ,\left\{\lambda_a\right\}\right) \, = \, \mathsf{a}(\mu)  \prod _{a=1}^N f\left(\mu ,\lambda _a\right)+ e^{i\vartheta}\mathsf{d}(\mu)  \prod _{a=1}^N f\left(\lambda _a,\mu \right) \, . \label{EvalueofTM}
\end{align}
Therefore, the eigenvalue $\theta \left(\mu ,\left\{\lambda_a\right\}\right)$ is expressed in terms of symmetric polynomials of $\lambda_a$. 
As discussed above, the eigenvalue of the transfer matrix is actually a generating function of mutually commuting conserved charges (or Hamiltonians). 
Accordingly, we can identify the expectation value of conserved charges of the XXX$_{1/2}$ spin chain model with the twisted GLSM correlators with appropriate coefficients.


\subsection{Correlation functions in the 3d $\mathcal{N}=2$ theory and the XXZ$_{1/2}$ spin chain model}
\label{ssec:2dXXZ}
We consider topologically twisted 3d $\mathcal{N}=2$ $U(N_c)$ gauge theories with an adjoint chiral multiplet $\Phi$ and $N_f$ chiral and anti-chiral multiplet $Q^{a}_{\ i}$, $\widetilde{Q}_{a}^{\ i}$, $a=1, \ldots, N_c$, $i=1, \ldots, N_f$, respectively, where we use the same notation as in the 2d case. 
There are flavor symmetries $SU(N_f)_Q$, $SU(N_f)_{\widetilde{Q}}$, and $U(1)_D$. 
In addition, there is a $U(1)_T$ topological symmetry in three dimensions. 
The matter contents and charge assignment are specified in Table \ref{table:3d}.

\begin{table}[h]
\begin{center}
\begin{tabular}{c | c c c c c c}
			&	$U(N_c)$			&	$SU(N_f)_{Q}$		&	$SU(N_f)_{\widetilde{Q}}$	&	$U(1)_D$	&	$U(1)_T$	&	$U(1)_R$	\\
			\hline
$Q$			&	$N_c$			&	$\overline{N}_f$			&	$\mathbf{1}$					&	$-1/2$	&	0			&	$r_1$		\\
$\widetilde{Q}$	&	$\overline{N}_c$	&	$\mathbf{1}$				&	$N_f$				&	$-1/2$	&	0			&	$r_2$		\\
$\Phi$		&	\text{adj}			&	$\mathbf{1}$				&	$\mathbf{1}$					&	$1$		&	0			&	$R$
\end{tabular}
 \caption{Matter contents of 3d $\mathcal{N}=2$ theory}
\label{table:3d}
\end{center}
\end{table}

We denote fugacities and magnetic fluxes of the Cartan part of global symmetries as follows;
\begin{align}
SU(N_f)_Q \, : \, (y_i, n_i)	\,	,	\quad 
SU(N_f)_{\widetilde{Q}} \, : \, (\widetilde{y}_i, \widetilde{n}_i)	\,	,	\quad 
U(1)_D \, : \, (z, l)	\,	,	\quad
U(1)_T \, : \, (\zeta, u)	\,	.
\end{align}
Then the topologically twisted index of the 3d $\mathcal{N}=2$ theory is given by 
\begin{align}
\begin{split}
&\hspace{-8mm} Z^\text{3d} \, = \, \frac{1}{N_c !} \sum_{\vec{m} \in \mathbb{Z}^{N_c}} \oint \prod_{a=1}^{N_c} \frac{d x_a}{2\pi i x_a}(-1)^{(N_c-1)\sum_{a=1}^{N_c} m_a} \prod_{a \neq b}^{N_c} \left( 1-  \frac{x_a}{x_b}\right) \prod_{a, b=1}^{N_c} \left( \frac{x_a^{1/2} x_b^{-1/2} z^{1/2} }{1-x_a x_b^{-1} z} \right)^{m_a-m_b+l-R+1} \\
&\hspace{30mm}  \times \prod_{a=1}^{N_c} \zeta^{m_a} \prod_{i=1}^{N_f}  
\left( \frac{x_a^{1/2} y_i^{-1/2} z^{-1/4}}{1 -x_a y^{-1}_i z^{-1/2}} \right)^{m_a - n_i - \frac{l}{2}- r_1+1} 
\left( \frac{x_a^{-1/2} \widetilde{y}_i^{1/2} z^{-1/4}}{1 -x_a^{-1} \widetilde{y}_i z^{-1/2}} \right)^{-m_a + \widetilde{n}_i - \frac{l}{2}- r_2+1}	\,	.
\label{3d-ttind}
\end{split}
\end{align}
Here $x_a$ is a constant value of the Wilson loop for the $a$-th diagonal $U(1)$ of the gauge group $U(N_c)$.
We take, for example, $\eta = (-1,-1, \ldots, -1)$ to choose a contour so that it picks poles from anti-chiral multiplets and the ajdoint chiral multiplet.
Poles exist when $m_a< \widetilde{n}_i - \frac{l}{2}- r_2+1$, and we resum over $m_i < K$ for a sufficiently large positive integer $K$. 
With $f_i = -n_i - \frac{l}{2} - r_1 + 1$, $\widetilde{f}_i =\widetilde{n}_i -\frac{l}{2} -r_1 +1$, and $h=l-R+1$, \eqref{3d-ttind} is written as
\begin{align}
\begin{split}
\hspace{-5mm} Z^\text{3d} \; =& \; \frac{ (-1)^{\frac{N_c(N_c+1)}{2}} z^{h N_c^2/2}}{N_c ! (1-z)^{hN_c}} 
\sum_{\vec{m} \in \mathbb{Z}^{N_c}} \oint \prod_{a=1}^{N_c} \frac{d x_a}{2 \pi i x_a} 
\prod_{b\neq a}^{N_c} \frac{1-\frac{x_a}{x_b}}{\left( 1- \frac{x_a}{x_b}z \right)^h}  
\prod_{a=1}^{N_c} \prod_{i=1}^{N_f} 
\frac{  (x_a y^{-1}_i z^{-1/2})^{f_i/2} (x_a^{-1} \widetilde{y}_i z^{-1/2})^{\widetilde{f}_i/2}}{(1-x_a y^{-1}_i z^{-1/2})^{f_i} 
(1-x_a^{-1} \widetilde{y}_i z^{-1/2})^{\widetilde{f}_i}} \\
& \qquad \qquad   \times 
\prod_{a=1}^{N_c }   \zeta^{m_a} (-1)^{(N_c-1)m_a} \prod_{b=1 \atop b \neq a}^{N_c} \left( \frac{x_a-x_b   z}{x_b - x_a  z} \right)^{m_a}
  \prod_{i=1}^{N_f} \prod_{a=1}^{N_c } \left(  
\frac{ x_a y^{-1/2}_i \widetilde{y}^{-1/2}_i}{(1-x_a y^{-1}_i z^{-1/2}) (1-x_a^{-1} \widetilde{y}_i z^{-1/2})^{-1}} \right)^{m_a}	\,	.
\label{3d-ttind1}
\end{split}
\end{align}
Summing over all fluxes for $m_i < K$ in \eqref{3d-ttind1}, we get
\begin{align}
\begin{split}
\hspace{-12mm} Z^{3d} \; =& \; \frac{(-1)^{\frac{N_c(N_c+1)}{2}} z^{h N_c^2/2}}{N_c ! (1-z)^{hN_c}}  \oint \prod_{a=1}^{N_c} \frac{d x_a}{2 \pi i x_a} 
\prod_{b\neq a}^{N_c}  \frac{1-\frac{x_a}{x_b}}{\left( 1- \frac{x_a}{x_b}z \right)^h}  \\
&\hspace{40mm} \times \prod_{a=1}^{N_c} \prod_{i=1}^{N_f} \left( \frac{ (x_a y^{-1}_i z^{-1/2})^{1/2}}{1-x_a y^{-1}_i z^{-1/2} }\right)^{f_i} \left( \frac{(x_a^{-1} \widetilde{y}_i z^{-1/2} )^{1/2}}{ 1-x_a^{-1} \widetilde{y}_i z^{-1/2}} \right)^{\widetilde{f}_i} 
\frac{(\zeta e^{i B_a(x)})^K}{\zeta e^{iB_a(x)}-1} 
\end{split}
\end{align}
where $B_a(x)$ is given by
\begin{align}
\exp (i B_a(x)) & \, := \, 
\prod_{b=1 \atop b \neq a}^{N_c} \left( \frac{x_a-x_b   z}{x_b - x_a  z} \right)
  \prod_{i=1}^{N_f}  \left(  
\frac{x_a y^{-1/2}_i \widetilde{y}^{-1/2}_i}{(1-x_a y^{-1}_i z^{-1/2}) (1-x_a^{-1} \widetilde{y}_i z^{-1/2})^{-1}} \right)	\, .
\end{align}
Due to $(\zeta e^{i B_a(x)})^K$ factor in the numerator with a sufficiently large $K$, poles at $x_a=0$, $1-x^{-1}_a \widetilde{y}_i z^{-1/2}=0$, and $x_a -z x_b=0$ are not available and only relevant poles come from $\zeta e^{iB_a(x)} =1$ for all $a$. 
We denote the solution for this equation by
\begin{align}
P_\text{3d} \, = \,  \{ (x_1, \cdots, x_{N_c}) \, | \,  \zeta e^{i B_a(x)} =1, \; \text{for all} \; a=1, 2, \ldots, N_c \}/S_{N_c}	\,	
\end{align}
where solutions that are related by Weyl permutations $S_{N_c}$ of $(x_1, \cdots, x_{N_c})$ are identified.
With
\begin{align}
x_a = e^{2 \lambda_a}	\,	,	\quad 
y_j = \widetilde{y}_j = e^{2 \nu_j}	\,	,	\quad 
z=e^{4i \eta}	\,	,	\quad	
(-1)^{N_f} \zeta=  e^{i \vartheta} \,	, 
\label{eq:3dparameters}
\end{align}
the contour integral becomes
\begin{align}
\begin{split}
Z^\text{3d} =& \frac{1}{(2i \sin 2\eta)^{N_c}} \, \sum_{\lambda_a \in P_\text{3d}} \, (\det  \mathcal{M}^{\text{3d}})^{-1} \prod_{a < b}^{N_c} (x_b x_a)^{h-1}
\prod_{a \neq b}^{N_c}\frac{ \sinh(\lambda_a - \lambda_b)}{\big( \sinh(\lambda_a - \lambda_b -2i\eta) \big)^h}\\
&\hspace{25mm} \times \left( \prod_{a=1}^{N_c} \prod_{i=1}^{N_f} \bigg( \frac{1}{2\sinh(\lambda_a - \nu_i - i \eta)} \bigg)^{f_i} \bigg( -\frac{1}{2\sinh(\lambda_a - \nu_i + i \eta)} \bigg)^{\widetilde{f}_i}\right) 
\label{par3dfinal}
\end{split}
\end{align}
where we used 
\begin{align}
\frac{\partial e^{iB_a(x)}}{\partial x_b} = \frac{ e^{iB_a(x)} }{x_b } \frac{ \partial B_a(x) }{\partial \lambda_b}
\end{align}
with
\begin{align}
\begin{split}
&\mathcal{M}^{\text{3d}}_{a b}:=- \frac{\partial i B_a(x)}{\partial \lambda_b} \, = \\   
&\hspace{-8mm} \delta_{ab} \left[   \sum_{j=1}^{N_f} \left( \frac{\cosh(\lambda_a - \nu_j - i \eta)}{\sinh(\lambda_a - \nu_j - i \eta)} - \frac{\cosh(\lambda_a - \nu_j + i \eta)}{\sinh(\lambda_a - \nu_j + i \eta)} \right) +  \sum_{e=1}^{N_c} \left( \frac{\cosh(\lambda_a - \lambda_e + 2i \eta)}{\sinh(\lambda_a - \lambda_e + 2i \eta)} - \frac{\cosh(\lambda_a - \lambda_e - 2i \eta)}{\sinh(\lambda_a - \lambda_e - 2i \eta)} \right) \right] \\
& \quad  -\left( \frac{\cosh(\lambda_a - \lambda_b + 2i \eta)}{\sinh(\lambda_a - \lambda_b + 2i \eta)} - \frac{\cosh(\lambda_a - \lambda_b - 2i \eta)}{\sinh(\lambda_a - \lambda_b - 2i \eta)} \right) \, .
\end{split}
\end{align}
Also, upon \eqref{eq:3dparameters}, the condition for supersymmetric vacua, $\zeta e^{iB_a(x)} =1$, \textit{i.e.} 
\begin{align}
\prod_{j=1}^{N_f} \frac{\sinh (\lambda_a - \nu_j - i \eta)}{\sinh (\lambda_a - \nu_j + i \eta)} \, = \, e^{i\vartheta} \prod_{b \neq a}^{N_c} \frac{\sinh(\lambda_b-\lambda_a+ 2 i \eta)}{\sinh(\lambda_b-\lambda_a - 2 i \eta)}
\end{align}
is exactly same as the Bethe ansatz for the XXZ$_{1/2}$ spin chain \eqref{BeqXXZ} with $N_c = N, \, N_f = M$.

If we choose $R$-charges and magnetic fluxes in such a way that
\begin{align}
r_1+n_i+\frac{l}{2} = 0	\,	,	\quad
r_2-\widetilde{n}_i+\frac{l}{2} = 0	\,	,	\quad 
R-l=0
\end{align}
hold, then the 3d topologically twisted index \eqref{par3dfinal} and the inverse of the norm of the Bethe eigenstate of the XXZ$_{1/2}$ spin chain model agree 
\begin{align}
Z^{3d} =\sum_{ (\lambda ) \in P_{\text{XXZ}}} \langle \Psi_N( \lambda )| \Psi_N( \lambda )\rangle^{-1} 
\end{align}
up to overall constants. \\

We can also consider correlation functions and conserved charges in the 3d $\mathcal{N}=2$ theory and the XXZ$_{1/2}$ spin chain model as in section \ref{ssec:GLSM1}. 
The eigenvalue $Q(u)$ of the Baxter $Q$-operator $\mathbf{Q}(u)$ in the XXZ$_{1/2}$ model is given by
\begin{align}
Q(u) \, = \, \prod_{a=1}^{N} \sinh(u - \lambda_a) \, ,
\end{align}
or $\frac{1}{2^N} \, e^{-Nu - \sum_{a=1}^{N} \lambda_a} \prod_{a=1}^{N}(e^{2u}-e^{2\lambda_a})$. Meanwhile, the Wilson loop in 3d $\mathcal{N}=2$ theories is given by the Schur polynomial
\begin{align}
W_{\mathcal{R}}(x) \, = \, s_{Y} (x_1, \ldots, x_{N_c} )
\end{align}
where $Y$ is the Young diagram for the representation $\mathcal{R}$ of $U(N_c)$. 
When $\mathcal{R}$ is a totally antisymmetric representation $Y \, = \, 1^r, \, r=1, \ldots, N_c$, the Schur polynomial is given by the elementary symmetric polynomials, $s_{1^r}(x_1, \ldots, x_{N_c}) \, = \, e_r(x_1, \ldots, x_{N_c})$. 
Therefore, with the identifications \eqref{eq:3dparameters}, the expectation value of Wilson loop operators is proportional to the coefficient of the eigenvalue of the Baxter $Q$-operator.

Also, as the eigenvalue of the transfer matrix $\tau(\mu)$ for the XXZ$_{1/2}$ model is given by \eqref{EvalueofTM} with \eqref{vacEvalXXZ}, we can identify the expectation value of conserved charges of the XXZ$_{1/2}$ model with the expectation value of Wilson loops with appropriate coefficients.


\section{Equivariant quantum cohomology, GLSM, and integrable model}
\label{sec:EQC}
In the previous section, we studied the relation between the A-twisted $\mathcal{N}=(2,2)$ GLSM and the XXX$_{1/2}$ spin chain. 
It was shown in \cite{Morrison:1994fr} that integrations of cohomology classes of toric Fano manifolds can be interpreted as correlation functions of $\sigma$ of the corresponding A-twisted $\mathcal{N}=(2,2)$ GLSM where the cup product of cohomology classes are deformed by using three point Gromov-Witten invariants (quantum cup product).
We may expect that such a relation holds for $\mathcal{N}=(4,4)$ GLSM where the target space is a hyperK\"ahler manifold.
We turn on all the possible twisted mass parameters including the one for the $\mathcal{N}=(2,2)$ adjoint chiral multiplet.\footnote{Since the $\mathcal{N}=(2,2)$ adjoint chiral multiplet with twisted mass parameter being turned off gives rise to the flat direction, we have to turn on the $U(1)_D$ twisted mass.}
In this section, we consider correlation functions of the A-twisted $\mathcal{N}=(2,2)^*$ GLSM on $S^2$ and study its relation to the equivariant quantum cohomology of the cotangent bundle of the Grassmannian.


\subsection{Equivariant quantum cohomology and equivariant integration}
Firstly, we summarize the equivariant quantum cohomology of the cotangent bundle of the Grassmannian $T^* \mathrm{Gr}(r, n)$ \cite{GRTV}.  
The Grassmannian $\mathrm{Gr}(r,n)$ is specified by the chains of subspaces,
\begin{align}
0 \, = \,  F_0 \, \subset \, F_1 \, \subset \, F_2  \, =  \, \mathbb{C}^{n}
\end{align}
with $\mathrm{dim} \, F_1=r$. 
We would like to consider the cotangent bundle $T^* \mathrm{Gr}(r,n)$ of the Grassmannian $\mathrm{Gr}(r,n)$.
We sometimes denote $(r, n-r)$ by $(\lambda_1, \lambda_2):=(r, n-r)$ below.

There is a torus action $(\mathbb{C}^*)^n \subset GL_n(\mathbb{C})$ on $\mathbb{C}^n$, accordingly on $\mathrm{Gr} (r,n)$. 
In addition, there is also a $\mathbb{C}^*$ action on the  fiber direction of $T^* \mathrm{Gr} (r,n)$. 
With these actions, one can consider a $GL_n(\mathbb{C}) \times \mathbb{C}^*$ equivariant cohomology ring.
The set of the Chern roots of bundles on $\mathrm{Gr}(r,n)$ with fiber $F_i/F_{i-1}$ is denoted by $\Gamma_i = \{ \gamma_{i,1}, \cdots, \gamma_{i, \lambda_i} \}$ with $i=1,2$.  
Also, the Chern root corresponding to each factors of $(\mathbb{C}^*)^n$ action and $\mathbb{C}^*$ action is denoted by $\mathbf{z}=\{ z_1; \cdots; z_n \}$ and $h$, respectively.
Then the equivariant cohomology ring is given by
\begin{align}
H^*_{GL_n(\mathbb{C}) \times \mathbb{C}^*} (T^* \mathrm{Gr} (r, n);\mathbb{C})
\, = \, \mathbb{C}[\mathbf{z}, \Gamma, h]^{S_{n} \times S_{\lambda_1} \times S_{\lambda_2} } \,  /  \, \mathcal{I}
 \label{equivCoh}
\end{align}
where $S_{n}$, $S_{\lambda_1}$ and $S_{\lambda_2}$ denote the symmetrization of variables $\{ z_1, \cdots, z_n \}$, $\{ \gamma_{1,1}, \cdots, \gamma_{1, \lambda_1} \}$ and $\{ \gamma_{2,1}, \cdots, \gamma_{2,\lambda_2} \}$, respectively.
The ideal $\mathcal{I}$ is generated by $n$ coefficients of a degree $n-1$ polynomial of $u$,
\begin{align}
\prod_{a=1}^{2} \prod_{b=1}^{\lambda_a} (u-\gamma_{a,b}) - \prod_{i=1}^{n} (u-z_i)	\, . 
\end{align}
The equivariant quantum cohomology ring of the cotangent bundle of the Grassmannian is given by
\begin{align}
QH^*_{GL_n(\mathbb{C}) \times \mathbb{C}^*} (T^* \mathrm{Gr} (r, n);\mathbb{C})
\, = \, \mathbb{C}[\mathbf{z}, \Gamma, h]^{S_{n} \times S_{\lambda_1} \times S_{\lambda_2} } \otimes \mathbb{C}[[{\sf q}]] \, / \, \mathcal{I}_{{\sf q}}
 \label{eq:QH}
\end{align}
where $\mathbb{C}[[{\sf q}]]$ is a ring of formal series of the quantum parameter ${\sf q}$. 
The ideal $\mathcal{I}_{{\sf q}}$ is generated by $n$ coefficients, $p_l$, defined by 
\begin{align}
\begin{split}
\hspace{-8mm} \sum_{l=1}^{n} p_{l} (\mathbf{z}, \Gamma, h, {\sf q}) u^{n-l}& \, := \,  
\prod_{a=1}^{2} \prod_{b=1}^{\lambda_a} (u-\gamma_{a,b}) \\
&\hspace{5mm} -{\sf q}  \prod_{a=1}^{\lambda_1} (u-\gamma_{1,a}-h) \prod_{b=1}^{\lambda_2} (u-\gamma_{2,b}+h) 
- (1-{\sf q})\prod_{i=1}^{n} (u-z_i) . \label{eq:ideal}
\end{split}
\end{align}
The coefficients $p_l$ are degree $l$ polynomials of each $\Gamma$ and $\mathbf{z}$, and are invariant under the action of $S_{n} \times S_{\lambda_1} \times S_{\lambda_2}$. 
Meanwhile, in \cite{Maulik:2012wi} the Yangian acting on the equivariant cohomology was constructed and the equivariant quantum cohomology ring was identified with the Bethe subalgebra of the integrable model.
The cotangent bundle of the Grassmannian is a typical example of \cite{Maulik:2012wi}.

The equivariant integration of the cohomology class $[f (\Gamma, \mathbf{z},h)] \in H^*_{GL_n(\mathbb{C}) \times \mathbb{C}^*} (T^* \mathrm{Gr} (r, n);\mathbb{C})$ is calculated by the formula
\begin{align}
\int_{T^* \mathrm{Gr}(r,n)} [f] \, = \, (-1)^{ \lambda_1 \lambda_2} \sum_{I_r \subset I} \prod_{i \in I_r} \prod_{j \in I_{n-r}} 
\frac{f(\mathbf{z}_{I},\mathbf{z};h)}{(z_i-z_j) (z_i-z_j+h) }
 \label{equivint}
\end{align}
where $I_r$ is a subset of $I=\{1, \cdots,n \}$ with $|I_r|=r$ and $I_{n-r}$ is the complement of $I_r$ in $I$. 
The factor $f(\mathbf{z}_{I},\mathbf{z};h)$ in the numerator is defined by the substitution $\Gamma=(\Gamma_1, \Gamma_2) \to (\mathbf{z}_{I_{r}},  \mathbf{z}_{I_{n-r}})$ in $f (\Gamma, \mathbf{z},h)$. 
Summation $\sum_{I_{r} \subset I}$ in \eqref{equivint} runs for all the possible subsets in $I$ with fixed $r$.

In section \ref{ssec:GLSM-EQC} we calculate the equivariant integration of the elements $[f (\Gamma, \mathbf{z},h;q)]$ in the equivariant quantum cohomology ring $QH^*_{GL_n(\mathbb{C}) \times \mathbb{C}^*} (T^* \mathrm{Gr} (r, n);\mathbb{C})$ for several examples by using the formula \eqref{equivint} and check that they match with the corresponding GLSM correlators.
More specifically, given a ring element, we reduce the degree of the ring element by using the ideal $\mathcal{I}_{{\sf q}}$ whenever it is possible and then apply the formula \eqref{equivint} to the resulting ring element, which depends on the parameter ${\sf q}$ in general.\footnote{As we will discuss, we expect that it works for $T^*\text{Gr}(r,n)$ when $r\leq n-r$ and we calculate differently when $r > n-r$.} 
The GLSM correlation function of the operator corresponding to a given original ring element before reducing is expected to match with the result of the equivariant integration obtained in a way we have just described.


\subsection{Correlation functions of A-twisted GLSM and equivariant integration of equivariant quantum cohomology}
\label{ssec:GLSM-EQC}
We study the relation between correlation functions of the A-twisted 2d $\mathcal{N}=(2,2)^*$ GLSM and the equivariant integration for the equivariant quantum cohomology classes in the cotangent bundle of the Grassmannian.

The gauge group and the matter contents are the same as in section \ref{ssec:GLSM1}, but we choose different $R$-charges from the previous case in such a way that we now have the superpotential
\begin{align}
W_{\widetilde{Q} \Phi Q} = \sum_{a,b=1}^{N_c} \sum_{i=1}^{N_f} \widetilde{Q}_a^{\ i} \Phi^{a}_{\ b} Q^{b}_{\ i}.
\end{align}
In the positive FI-parameter region, the target space of the non-linear sigma model limit of the theory is $T^{*} \mathrm{Gr}(N_c,N_f)$ where the base space $\mathrm{Gr}(N_c,N_f)$ is parametrized by $Q^{b}_{\ i}$. 
On the other hand, in the negative FI-parameter region, the target is again $T^{*} \mathrm{Gr}(N_c,N_f)$ but the base space is parametrized by $\widetilde{Q}_a^{\ i}$.

The superpotential breaks $SU(N_f)_Q \times SU(N_f)_{\widetilde{Q}} \times U(1)_D$ to $SU(N_f) \times U(1)_D$.
We turn off all the background fluxes for flavor symmetry groups.
The twisted mass parameters for the $SU(N_f)$ flavor symmetry are denoted by $m_i$ and the twisted mass parameter for $U(1)_D$ flavor symmetry by $m^z$.   

\begin{table}[h]
\begin{center}
\begin{tabular}{c |  c c c c }
			&	$U(N_c)$			&	$SU(N_f)$			&	$U(1)_D$		&	$U(1)_R$	\\
			\hline
$Q$			&	$N_c$			&	$\overline{N}_f$								&	$-1/2$				&	$0$		\\
$\widetilde{Q}$	&	$\overline{N}_c$					&	$N_f$				&	$-1/2$				&	$0$		\\
$\Phi$		&	\text{adj}			&	$\mathbf{1}$									&	$1$					&	$2$
\end{tabular} \caption{The charge assignment of GLSM}
\label{Table:GLSMq}
\end{center}
\end{table}

The correlation function of the gauge invariant operator $\mathcal{O}(\sigma)$ constructed from $\sigma=\mathrm{diag} (\sigma_1, \cdots, \sigma_{N_c})$ is  
\begin{align}
\begin{split}
\langle \mathcal{O}(\sigma) \rangle^{N_c,N_f}_{\rm A \mathchar `-twist} \, =& \, \frac{(-1)^{N_*}}{N_c !}\sum_{a=1}^{N_c}\sum_{k_a=0}^{\infty} ((-1)^{N_c-1}q)^{\sum_{a=1}^{N_c} k_a} \,
\oint \prod_{a=1}^{N_c} \frac{d \sigma_a}{2\pi i} \,  \mathcal{O}(\sigma) \\
&\hspace{8mm}\times \frac{\prod_{1 \le a \neq b \le N_c}(\sigma_a-\sigma_b)}{\prod_{a,b=1}^{N_c} (\sigma_a-\sigma_b+m^z)^{k_a-k_b-1}} \,
\prod_{i=1}^{N_f}\prod_{a=1}^{N_c} \frac{(-\sigma_a+m_i-\frac{m^z}{2})^{k_a-1}}{(\sigma_a-m_i-\frac{m^z}{2})^{k_a+1}} .
\label{eq:Atwistcor} 
\end{split}
\end{align}
Here we take the charge vector in the Jeffrey-Kirwan reisdue formula as $\mathrm{Re} \, q <1$.
Then residues are evaluated at  the poles $(\sigma_a-m_i-\frac{m^z}{2})^{-(k_a+1)}$ and it is easy to show that poles coming from $(\sigma_a-\sigma_b+m^z)^{-(k_a-k_b-1)}$ do not contribute to the residues.
The overall sign ambiguity will be fixed below.


\subsubsection*{Bethe ansatz equation and matching of parameters}
Before considering correlation functions, let us see how the twisted chiral ring relation \eqref{preBetheeq} and the Bethe ansatz equation, $\exp \left(2 \pi i \frac{\partial \widetilde{\mathcal{W}}_{\mathrm{eff}}}{\partial \sigma_a}\right) =1$, arise from the ideal of the equivariant quantum cohomology \eqref{eq:ideal} to identify the parameters.

By substituting  $u=\gamma_{1,c}$ and $\gamma_{1,c}+h$ into 
\begin{align}
0 \, = \, \sum_{l=1}^{n} \, p_{l}  (\mathbf{z}, \Gamma, h, {\sf q}) \, u^{n-l} \, ,  
\label{eq:qidealzero}
\end{align}
we obtain two equations
\begin{align}
 -{\sf q}  \prod_{a=1}^{\lambda_1} (\gamma_{1,c}-\gamma_{1,a}-h) \prod_{b=1}^{\lambda_2} (\gamma_{1,c}-\gamma_{2,b}+h)
\, &= \,  (1-{\sf q})\prod_{i=1}^{n} (\gamma_{1,c}-z_i)	\,	, 
\label{eq:QQ1}
\\
\prod_{a=1}^{\lambda_1} (\gamma_{1,c}-\gamma_{1,a}+h) \prod_{b=1}^{\lambda_2} (\gamma_{1,c}-\gamma_{2,b}+h) 
\, &= \, (1-{\sf q})\prod_{i=1}^{n} (\gamma_{1,c}-z_i+h) 	\,	.
\label{eq:QQ2}
\end{align}
Dividing \eqref{eq:QQ1}  by \eqref{eq:QQ2}, we get 
\begin{align}
 {\sf q}  \prod_{a=1\atop a \neq c}^{\lambda_1} \frac{\gamma_{1,c}-\gamma_{1,a}+h}{\gamma_{1,c}-\gamma_{1,a}-h} 
=  \prod_{i=1}^{n} \frac{\gamma_{1,c}-z_i}{\gamma_{1,c}-z_i+h}		\,	 .
\label{eq:Bethelike}
\end{align}
Upon the identifications
\begin{align}
 r=N_c, ~~ n=N_f, ~~ \gamma_{1,a}=\sigma_a, ~~ z_i=m_i+\frac{m^z}{2}, ~~ h=m^z,~~ {\sf q}=(-1)^{N_f}q	\,	,
\label{eq:parameters1}
\end{align}
\eqref{eq:Bethelike} agrees with the condition for the supersymmetric vacua \eqref{preBetheeq} for the $\mathcal{N}=(2,2)^*$ $U(N_c)$ GLSM with $N_f$ fundamental hypermultiplets.


\subsubsection*{Quantum cohomology of $\mathbb{CP}^{n-1}$ and correlation functions of the A-twisted GLSM}
We briefly recall the well-known relation between the $\mathcal{N}=(2,2)$ $U(1)$ GLSM with $n$ charge $+1$ chiral multiplets and the quantum cohomology of $\mathbb{CP}^{n-1}$. 
This GLSM flows to the $\mathcal{N}=(2,2)$ non-linear sigma model with target space $\mathbb{CP}^{n-1}$ \cite{Witten:1993yc, Morrison:1994fr}.

The quantum cohomology of $\mathbb{CP}^{n-1}$ is given by 
\begin{align}
QH^* (\mathbb{CP}^{n-1} ; \mathbb{C}) \, \simeq \, \mathbb{C}[{\sf q}, \gamma_{1,1}] \, / \, (\gamma^n_{1,1} -{\sf q}) .
\end{align}
The equivariant integration of $\gamma^{l}_{1,1} \in QH^* (\mathbb{CP}^{n-1} ; \mathbb{C})$, which we denote as $\langle \gamma^{l}_{1,1} \rangle_{\mathbb{CP}^{n-1}}$,
is obtained as follows.
If $a < n$, $\langle \gamma^a_{1,1} \rangle_{\mathbb{CP}^{n-1}}$ is the same as the integral of the cohomology class $\gamma^a_{1,1} \in H^* (\mathbb{CP}^{n-1} ; \mathbb{C})$ and is given by
\begin{align}
\langle \gamma^a_{1,1} \rangle_{\mathbb{CP}^{n-1}}= \int_{\mathbb{CP}^{n-1}} \gamma^a_{1,1} \, = \,
\begin{cases}
\, 1 & a=n-1 \, \\
\,  0 & a < n-1 \, .
\end{cases}
\end{align}
For $\langle \gamma^{ m n +a }_{1,1} \rangle_{\mathbb{CP}^{n-1}}$ with $a<n$, we reduce the degree by using the relation $\gamma^n_{1,1} -{\sf q}=0$ to $\gamma^{ m n +a }_{1,1}={\sf q}{}^m \gamma^{a}_{1,1}$ and obtain
\begin{align}
\langle \gamma^{ mn+a }_{1,1} \rangle_{\mathbb{CP}^{n-1}} \, = \, {\sf q}{}^{m} \, \int_{\mathbb{CP}^{n-1}} \gamma^a_{1,1} \, = \,
\begin{cases}
\, {\sf q}{}^m & a=n-1 \\
\, 0 & a < n-1	\,	.
\end{cases}
\end{align}
On the other hand, the expectation value of {$\sigma^{l}$ is obtained by supersymmetric localization     
\begin{align}
\langle \sigma^{l} \rangle_{\rm A \mathchar `-twist} \, = \,  \sum_{k=0}^{\infty} q^{ k} 
\oint _{\sigma=0} \frac{d \sigma}{2\pi i}  \, \sigma^{l-n(k+1)}  \, ,
\end{align}
which gives
\begin{align}
\langle \sigma^{mn + a} \rangle_{\rm A \mathchar `-twist} \, = \, 
\begin{cases}
\, q^{m} & a=n-1 \\
\, 0 & a < n-1 \, .
\end{cases}
\end{align}
Therefore we have 
\begin{align}
\langle \gamma^{l}_{1,1} \rangle_{\mathbb{CP}^{n-1}} \, = \, \langle \sigma^{l} \rangle_{\rm A \mathchar `-twist}, \; \; {\rm with}  \; q={\sf q} \, .
\label{eq:massid}
\end{align} 
We perform a similar calculation for the cotangent bundle of the Grassmannian.


\subsubsection{{$T^*\mathbb{CP}^{n-1}$}}

We would like to relate the expectation value of $\sigma^l$ in the GLSM to the equivariant integration of equivariant quantum cohomology classes when the target space is $T^*\mathbb{CP}^{n-1}$.


\subsubsection*{\textbullet \ $T^*\mathbb{CP}^1$}
When $\lambda_1=1$ and $n=2$, \eqref{eq:Bethelike} becomes
\begin{align}
\gamma^2_{1,1} \, = \, (z_1+z_2)\gamma_{1,1} + \frac{2h {\sf q} }{1-{\sf q}} \gamma_{1,1} +\frac{h {\sf q} (h- z_1- z_2)+{\sf q} z_1 z_2}{1-{\sf q}}  \, .
\label{eq:chiralCP1}
\end{align}
This relation is the same as the twisted chiral ring relation via \eqref{eq:parameters1}.
By using \eqref{eq:chiralCP1}, $\gamma^l_{1,1}$ can be uniquely expressed as
\begin{align}
\gamma^l_{1,1}=A^{(1)}_l(z, h,{\sf q}) \gamma_{1,1}+A^{(0)}_{l} (z, h,{\sf q}) \,	.
\label{eq:reduceCP1} 
\end{align}
From \eqref{eq:reduceCP1}, the equivariant integration of $\gamma^{l}_{1,1}$ is given by 
\begin{align}
\langle \gamma^l_{1,1} \rangle_{T^*\mathbb{CP}^1}
= A^{(1)}_l(z, h,{\sf q}) \int_{T^*\mathbb{CP}^1} [ \gamma_{1,1} ]+A^{(0)}_{l} (z, h,{\sf q}) \int_{T^*\mathbb{CP}^1} [1]  \, .
\label{eq:QintCP1}
\end{align}
The correlation function $\langle \sigma^l \rangle^{N_c=1,N_f=2}_{\rm A \mathchar `-twist}$ is expected to be related to the equivariant integral on $T^*\mathbb{CP}^1$
\begin{align}
\langle \sigma^l \rangle^{N_c=1, N_f=2}_{\rm A \mathchar `-twist}= \langle \gamma^l_{1,1} \rangle_{T^*\mathbb{CP}^1} 
\label{eq:GLSMQM}
\end{align}
via the identification of parameters \eqref{eq:parameters1}.
We can check this explicitly. 
For example, when $l \leq 1$, the equivariant integration $\langle \gamma^l_{1,1} \rangle_{T^*\mathbb{CP}^1}$ gives
\begin{align}
\int_{T^*\mathbb{CP}^1} [1 ]
= \sum_{j=1}^2 \prod_{i=1 \atop i \neq j}^{2} \frac{-1}{(z_j-z_i+h)(z_j-z_i)} \, , \\
\int_{T^*\mathbb{CP}^1} [ \gamma_{1,1} ]
= \sum_{j=1}^2 \prod_{i=1 \atop i \neq j}^{2} \frac{-z_j}{(z_j-z_i+h)(z_j-z_i)} \, .
\end{align}
Meanwhile, we evaluate $\langle 1 \rangle^{N_c=1, N_f=2}_{\rm A \mathchar `-twist}$ for several orders of $q$ and see that there are no $q$ corrections,
\begin{align}
\langle 1 \rangle^{N_c=1, N_f=2}_{\rm A \mathchar `-twist} 
= \sum_{j=1}^2 \prod_{i=1 \atop i \neq j}^{2} \frac{-1}{(z_i-z_j)(z_i-z_j+h)} \, , \\
\langle \sigma \rangle^{N_c=1, N_f=2}_{\rm A \mathchar `-twist} 
= \sum_{j=1}^2 \prod_{i=1 \atop i \neq j}^{2} \frac{-z_j}{(z_i-z_j)(z_i-z_j+h)} \, .
\end{align}
Here we fixed the overall sign in order to have an agreement with the equivariant integration $\int_{T^*\mathbb{CP}^1} [1 ]$.
Therefore, we checked that
\begin{align}
\langle 1 \rangle^{N_c=1, N_f=2}_{\rm A \mathchar `-twist} = \langle 1 \rangle_{T^*\mathbb{CP}^1}	\, , \\   
\langle \sigma \rangle^{N_c=1, N_f=2}_{\rm A \mathchar `-twist} = \langle \gamma_{1,1} \rangle_{T^*\mathbb{CP}^1} \, .
\end{align}
We also computed $\langle \sigma^l \rangle^{N_c=1, N_f=2}_{\rm A \mathchar `-twist}$ perturbatively and $\langle \gamma^l_{1,1} \rangle_{T^*\mathbb{CP}^1}$ exactly by using \eqref{eq:QintCP1} for $l=2,3,4,5$, and checked agreement \eqref{eq:GLSMQM}.


\subsubsection*{\textbullet \ $T^*\mathbb{CP}^{n-1}$}
We expect that the expectation value of $\sigma^l$ agrees with the integration of 
$\gamma^l_{1,1} \in QH_{GL_n(\mathbb{C}) \times \mathbb{C}^*} (T^*\mathbb{CP}^{n-1};\mathbb{C})$,
\begin{align}
\langle \sigma^l \rangle^{N_c=1, N_f=n}_{\rm A \mathchar `-twist} = \langle \gamma^l_{1,1} \rangle_{T^*\mathbb{CP}^{n-1}}	\, .
\end{align}
From the ideal, we obtain the following relation 
\begin{align}
\prod_{i=1}^{n} (\gamma_{1,c}-z_i+h )+ {\sf q} \prod_{i=1}^{n}  (\gamma_{1,c}-z_i)=0 .
\label{eq:chiralCP2}
\end{align}
This relation is the same as the twisted chiral ring relation of the corresponding GLSM via \eqref{eq:parameters1}.
From \eqref{eq:chiralCP2},  $\gamma^l_{1,1}$ with $l >n-1$ is uniquely expressed as
\begin{align}
 \gamma^l_{1,1} \, = \, \sum_{k=0}^{n-1} A^{(k)}_l(z, h,{\sf q}) \, \gamma^k_{1,1} \, .
\end{align}
With the identification $\sigma=\gamma_{1,1}$, we expect that $\langle \sigma^l \rangle^{N_c=1,N_f=n}_{\rm A \mathchar `-twist}$ agrees with the equivariant integration of the equivariant quantum cohomology class $\gamma^l_{1,1}$,
\begin{align}
\langle \sigma^l \rangle^{N_c=1, N_f=n}_{\rm A \mathchar `-twist}
= \sum_{k=0}^{n-1} A^{(k)}_l(z, h, {\sf q}) \int_{T^* \mathbb{CP}^{n-1}} [\gamma^k_{1,1}] .
\end{align}
We also checked this for $n=3,4$ with several higher powers of $\sigma$ and found agreement.


\subsubsection{$T^*\mathrm{Gr}(r,n)$ with $r \le n-r$}
\label{corrGR}
We consider the non-Abelian cases with $r \le n-r$. 
The case with $r  > n-r $ will be discussed in section \ref{seiberg}. 
As an example, we consider $T^*\text{Gr}(2,4)$. 
The general symmetric polynomial takes the form of $(\gamma_{1,1} + \gamma_{1,2})^k(\gamma_{1,1} \gamma_{1,2})^l$ where $k$ and $l$ are non-negative integers. 
When $k+l \geq3$, $(\gamma_{1,1} + \gamma_{1,2})^k(\gamma_{1,1} \gamma_{1,2})^l$ can be reduced to $(\gamma_{1,1} + \gamma_{1,2})^k(\gamma_{1,1} \gamma_{1,2})^l$ with $k+l \le 2$ by using the ideal.
Thus, for $k+l \le 2$, we expect
\begin{align}
\langle (\sigma_1+\sigma_2)^k  (\sigma_1 \sigma_2)^l \rangle^{N_c=2, N_f=4}_{\rm A \mathchar `-twist}
=\int_{T^*\mathrm{Gr}(2,4)} [ (\gamma_{1,1}+\gamma_{1,2})^k (\gamma_{1,1}\gamma_{1,2})^l ] \, .
\label{gr24leq2}
\end{align}
In other words, $\langle (\sigma_1+\sigma_2)^k  (\sigma_1 \sigma_{2})^l \rangle^{N_c=2, N_f=4}_{\rm A \mathchar `-twist}$ with $k+l \le 2$ would not have $q$ dependence.
On the other hand, when $k+l \geq 3$, we reduce the degree of $(\gamma_{1,1}+\gamma_{1,2})^k (\gamma_{1,1}\gamma_{1,2})^l$ by using the ideal,
\begin{align}
(\gamma_{1,1}+\gamma_{1,2})^k (\gamma_{1,1}\gamma_{1,2})^l
=\sum_{s+t \le 2} A^{(s,t)}_{k,l} (z,h,{\sf q})  (\gamma_{1,1}+\gamma_{1,2})^s (\gamma_{1,1}\gamma_{1,2})^t .
\label{eq:Gr24red}
\end{align} 
Therefore, we expect
\begin{align}
\langle (\sigma_1+\sigma_2)^k  (\sigma_1 \sigma_2)^l \rangle^{N_c=2, N_f=4}_{\rm A \mathchar `-twist}
=\sum_{s+t \le 2} A^{(s,t)}_{k,l} (z,h,{\sf q}) \int_{T^* \mathrm{Gr}(2,4)}  [(\gamma_{1,1}+\gamma_{1,2})^s (\gamma_{1,1}\gamma_{1,2})^t] .
\end{align} 
We have checked \eqref{gr24leq2} and \eqref{eq:Gr24red} for $k+l \leq 3$ perturbatively. 
The detailed calculation of the reduction \eqref{eq:Gr24red} is available in Appendix \ref{appA} as an example. 
We also checked the cases of $k+l=4$ and some of $k+l=5$ for $T^*\text{Gr}(2,5)$ and found agreement.
We expect to have agreement for general $r \leq n-r$.


\subsubsection{$T^*\mathrm{Gr}(r,n)$ with $r > n-r$ and the Seiberg-like duality}
\label{seiberg}
From the ideal $p_1=0$ for the equivariant quantum cohomology of $T^*\mathrm{Gr}(r,n)$, we have
\begin{align}
 \sum_{a=1}^{\lambda_1} \gamma_{1,a} + \sum_{a=1}^{\lambda_2} \gamma_{2,a}
=\frac{(\lambda_1-\lambda_2) {\sf q} h}{(1-{\sf q})} +\sum_{i=1}^n z_i 	\,	
\label{eq:chiralring1}
\end{align}
where $\lambda_1=r$ and $\lambda_2=n-r$.
In section \ref{corrGR}, we expected, for example,
\begin{align}
\Big\langle \sum_{a=1}^r  \sigma_a \Big\rangle^{N_c=r, N_f=n}_{\rm A \mathchar `-twist} =\int_{T^* \mathrm{Gr}(r,n)} \sum_{a=1}^r \gamma_{1,a} \, \quad {\rm for} \  r \le n-r \, ,  
\label{eq:QtoC}
\end{align}
\textit{i.e.} $\big\langle \sum_{a=1}^r \gamma_{1,a} \big\rangle_{T^* \mathrm{Gr}(r,n)} $ does not have any $q$ corrections and can be computed by using \eqref{equivint}.
From the relation \eqref{eq:chiralring1}, it is expected that $\big\langle \sum_{a=1}^r \gamma_{1,a} \big\rangle_{T^* \mathrm{Gr}(r,n)} $ with $r > n-r$ receives $q$ corrections and differs from the result directly obtained by the classical equivariant integration \eqref{equivint}. 
In order to calculate the equivariant integration properly for the case $r > n-r$, it is useful to study the isomorphism $T^* \text{Gr}(r,n) \simeq T^* \text{Gr}(n-r, n)$, which corresponds to the Seiberg-like duality \cite{Benini:2014mia} between A-twisted $\mathcal{N}=(2,2)^*$ GLSM's with gauge groups $U(r)$ and $U(n-r)$.

For this purpose, we consider the relation between ideals of the equivariant quantum cohomology of $T^* \mathrm{Gr} (r, n)$ and $T^* \mathrm{Gr} (n-r, n)$. The latter is given by 
\begin{align}
QH^*_{GL_n(\mathbb{C}) \times \mathbb{C}^*} (T^* \mathrm{Gr} (n-r, n);\mathbb{C})
\, \simeq \, \mathbb{C}[\tilde{\mathbf{z}}, \tilde{\Gamma}, \tilde{h}]^{S_{n} \times S_{\tilde{\lambda}_1} \times S_{\tilde{\lambda}_2} } \otimes \mathbb{C}[[\tilde{{\sf q}}]] \, / \, \tilde{\mathcal{I}}_{{\sf q}}
 \label{QH2}
\end{align}
where we use tilde to distinguish the notations for $QH^*_{GL_n(\mathbb{C}) \times \mathbb{C}^*} (T^* \mathrm{Gr} (r, n);\mathbb{C})$.
The ideal $\tilde{\mathcal{I}}^{\sf q}$ is generated by $n$ polynomials $\tilde{p}_l$ defined by
\begin{align}
\begin{split}
\sum_{l=1}^{n} \tilde{p}_{l} (\tilde{\mathbf{z}}, \tilde{\Gamma}, \tilde{h}, \tilde{{\sf q}}) u^{n-l} \, 
&:= \, \prod_{a=1}^{2} \prod_{b=1}^{\tilde{\lambda}_a} (u-\tilde{\gamma}_{a,b}) \\
&\hspace{3mm} -\tilde{{\sf q}}  \prod_{a=1}^{\tilde{\lambda}_1} (u-\tilde{\gamma}_{1,a}-h) \prod_{b=1}^{\tilde{\lambda}_2} (u-\tilde{\gamma}_{2,b}+\tilde{h})
- (1-\tilde{{\sf q}})\prod_{i=1}^{n} (u-\tilde{z}_i) \, .
\end{split}
\end{align}
The ideals of the quantum cohomology of $T^*\mathrm{Gr}(r,n)$ and of $T^*\mathrm{Gr}(n-r,n)$ are the same upon the following parameter identification\footnote{This identification was also discussed in \cite{Benini:2014mia}.}
\begin{align}
\gamma_{1,a}=\tilde{\gamma}_{2,a}-h\, , ~~  \gamma_{2,a}=\tilde{\gamma}_{1,a}+h \, , ~~ z_{i}=\tilde{z}_{i} \, , ~~  h=\tilde{h} \, ,  ~~ {\sf q}=\tilde{{\sf q}}{}^{-1} \, . 
\label{eq:generators}
\end{align}
When equivariant parameters are turned off, $\gamma_{1,a}$ and $\tilde{\gamma}_{2,a}$ are exchanged with each other under  $T^*\mathrm{Gr}(r,n) \leftrightarrow T^*\mathrm{Gr}(n-r,n)$.
This is consistent with  the fact that vector bundles with fibers  $F_1$  and $ F_2/F_1$ are exchanged vice versa under  $\mathrm{Gr}(r,n) \leftrightarrow \mathrm{Gr}(n-r,n)$.

Next we identify the variables in $QH^*_{GL_n(\mathbb{C}) \times \mathbb{C}^*} (T^* \mathrm{Gr} (n-r, n);\mathbb{C})$ with those in $U(n-r)$ GLSM. 
By substituting $u=\tilde{\gamma}_{1,c}$ and $\tilde{\gamma}_{1,c}+h$ into $\sum_{l=1}^{n} \tilde{p}_{l} (\tilde{\mathbf{z}}, \tilde{\Gamma}, \tilde{h}, \tilde{{\sf q}}) u^{n-l}=0$, we obtain 
\begin{align}
 \tilde{{\sf q}}  \prod_{a=1\atop a \neq c}^{\tilde{\lambda}_1} \frac{\tilde{\gamma}_{1,c}-\tilde{\gamma}_{1,a}+\tilde{h}}{\tilde{\gamma}_{1,c}-\tilde{\gamma}_{1,a}-\tilde{h}} 
=  \prod_{i=1}^{n} \frac{\tilde{\gamma}_{1,c}-\tilde{z}_i}{\tilde{\gamma}_{1,c}-\tilde{z}_i+\tilde{h}}	\,	 .
\label{eq:Bethelikedual}
\end{align}
With the identifications
\begin{align}
 n-r=N_c, ~~ n=N_f, ~~ \tilde{\gamma}_{1,a}=\tilde{\sigma}_a, ~~ z_i=\tilde{m}_i+\frac{\tilde{m}^z}{2}, ~~ \tilde{h}=\tilde{m}^z,~~ \tilde{{\sf q}}=(-1)^{N_f} \tilde{q}	\,	,
\label{eq:paradual}
\end{align}
\eqref{eq:Bethelikedual} also agrees with the twisted chiral ring relation.\footnote{There is another way of identification, but considering the Seiberg-like duality above identification is more appropriate.}
From \eqref{eq:generators}, \eqref{eq:parameters1} and \eqref{eq:paradual}, the relation of twisted mass parameters and FI-parameter in $U(r)$ and $U(n-r)$ GLSM is
\begin{align}
m_i = \tilde{m}_i, \quad  m^z=\tilde{m}^z,  ~~ q=\tilde{q}^{-1}	\,	. 
\label{eq:parameters3}
\end{align}	
\vspace{0mm}

We begin with the simplest case, which corresponds to the partition function.
From \eqref{equivint}, we obtain
\begin{align}
\int_{T^*\mathrm{Gr}(r,n)} [ 1 ] = \int_{T^*\mathrm{Gr}(n-r,n)}  [ 1 ] 
\label{eq:seiberg1}
\end{align}
and this implies
\begin{align}
\langle 1 \rangle^{N_c=r, N_f=n}_{\rm A \mathchar `-twist}=\langle 1 \rangle^{N_c=n-r, N_f=n}_{\rm A \mathchar `-twist}=\int_{T^*\mathrm{Gr}(r,n)} [ 1 ] 	\,	.
\label{eq:seibergPF}
\end{align}
We computed each side of \eqref{eq:seibergPF} for $(N_c,N_f)=(1,3), (1,4), (2,5)$ in several orders of $q$ and checked the agreement. \\

Next, with the identification \eqref{eq:generators}, we have
\begin{align}
\begin{split}
  \sum_{a=1}^{r} {\gamma}_{1,a}  \, & = \, 
 - \sum_{a=1}^{n-r} {\gamma}_{2,a}  +  \frac{(2r-n) {\sf q} {h}}{(1-{\sf q})} + \sum_{i=1}^n z_i   \\
 &= - \sum_{a=1}^{n-r} (\tilde{\gamma}_{1,a} +h) + \frac{(2r-n) {\sf q} {h}}{(1-{\sf q})} + \sum_{i=1}^n z_i \, . 
\label{eq:dualsigma1}
\end{split}
\end{align}
This gives a map between $\sigma$ in $U(r)$ and $\tilde{\sigma}$ in $U(n-r)$ GLSM's
\begin{align}
\begin{split}
  \sum_{a=1}^{r} \sigma_{a}  \, & = \, 
 - \sum_{a=1}^{n-r} (\tilde{\sigma}_{a} +h) + \frac{ (2r-n) {\sf q} {h}}{(1-{\sf q})} + \sum_{i=1}^n z_i \, . 
\label{eq:dualsigma2}
\end{split}
\end{align}
Therefore we get
\begin{align}
\begin{split}
  \Big\langle  \sum_{a=1}^{r} \sigma_{a} \Big\rangle^{N_c=r, \, N_f=n}_{\rm A \mathchar `-twist}  \, & = \, 
     \Big\langle - \sum_{a=1}^{n-r} (\tilde{\sigma}_{a} +h) + \frac{(2r-n) {\sf q} {h}}{(1-{\sf q})} + \sum_{i=1}^{n} z_i  \Big\rangle^{N_c=n-r, \, N_f=n}_{\rm A \mathchar `-twist} \, . 
\label{eq:dualsigma2}
\end{split}
\end{align}
We evaluate the LHS of \eqref{eq:dualsigma2} in the region $|q| <1$.  
From $q=\tilde{q}^{-1}$, the RHS is evaluated in the region $ |\tilde{q} | >1$ where the one-loop determinant of anti-chiral multiplets contribute to Jeffrey-Kirwan residues.
Note that the expression of $\langle  \sum_{a=1}^{n-r} (\tilde{\sigma}_{a} +h) \rangle^{N_c=n-r, \, N_f=n}_{\rm A \mathchar `-twist} $ with $| \tilde{q} | >1$ is the same as the expression of $\langle  \sum_{a=1}^{n-r} \tilde{\sigma}_{a}  \rangle^{N_c=n-r, \, N_f=n}_{\rm A \mathchar `-twist} $ with $| \tilde{q} | <1$ upon the change of parameter $\tilde{q}  \to \tilde{q}^{-1}$.
Since $\langle  \sum_{a=1}^{n-r} \tilde{\sigma}_{a}  \rangle^{N_c=n-r, \, N_f=n}_{\rm A \mathchar `-twist} $ for $r  >n-r$ does not depend on $\tilde{q}$ in \eqref{eq:QtoC}, we have
\begin{align}
\begin{split}
& \Big\langle \sum_{a=1}^{n-r} (\tilde{\sigma}_{a} +h) + \frac{(2r-n) {\sf q} {h}}{(1-{\sf q})} + \sum_{i=1}^{n} z_i   \Big\rangle^{N_c=n-r, \, N_f=n}_{\rm A \mathchar `-twist}   \\ 
\, & = \, 
     \sum_{I_{n-r} \subset I} \prod_{a \in I_{n-r}} \prod_{b \in I_{r}} \frac{ (-1)^{ r(n-r)}}{(z_a-z_b) (z_a-z_b+h) } \left(\sum_{c \in I_{r} }z_c +  \frac{(2r-n){\sf q} h}{(1-{\sf q})} +\sum_{i=1}^{n} z_i  \right) \, , ~~ 
\end{split}  
\label{eq:dualgamma3}
\end{align}
where $r > n-r$ and $|\tilde{q}| >1 $.
On the other hand, the correspondence between the A-twisted GLSM and the quantum cohomology provides 
\begin{align}
 \Big\langle \sum_{a=1}^{r} {\gamma}_{1,a} \Big\rangle_{T^*\mathrm{Gr}(r,n)}
= \Big\langle  \sum_{a=1}^{N_c } {\sigma}_{a} \Big\rangle^{N_c=r, \, N_f=n}_{\rm A \mathchar `-twist} \, .
\label{eq:intgamma2}
\end{align}
Therefore from \eqref{eq:dualsigma1} and \eqref{eq:intgamma2} we obtain 
\begin{align}
\Big\langle \sum_{a=1}^{r} {\gamma}_{1,a} \Big\rangle_{T^*\mathrm{Gr}(r,n)}
  & =\Big\langle  \sum_{a=1}^{N_c } {\sigma}_{a} \Big\rangle^{N_c=r, \, N_f=n}_{\rm A \mathchar `-twist}  \\
\, & = \, 
     \sum_{I_{n-r} \subset I} \prod_{a \in I_{n-r}} \prod_{b \in I_{r}} 
\frac{ (-1)^{ r(n-r)}}{(z_a-z_b) (z_a-z_b+h) } \left(\sum_{c \in I_{r} } z_c +  \frac{(2r-n){\sf q} h}{(1-{\sf q})} +\sum_{i=1}^{n} z_i \right) .
\label{eq:dualgamma3}
\end{align}
We computed $ \Big\langle  \sum_{a=1}^{r } {\sigma}_{a} \Big\rangle^{N_c=r, \, N_f=n}_{\rm A \mathchar `-twist}$ with $ |q| <1$ for $(N_c,N_f)=(2,3), (2,4), (3,4)$ in several orders of $q$ and checked the equality. \\

We can do similar calculations for other cases. 
For example, we consider $\mathrm{Tr}_{A_2} \sigma$.
Here, trace is taken over the second anti-symmetric representation. 
To make the expression concise, we call $e_{l} (\gamma_1), e_{l} (\tilde{\gamma}_1)$, $e_{l} (\gamma_2), e_{l} (\tilde{\gamma}_2)$, $e_{l} ({\sigma}),  e_{l} (\tilde{\sigma})$ and $e_{l} (z)$ the $l$-th elementary symmetric polynomials of $\gamma_{1,a}, \tilde{\gamma}_{1,a}$, $\gamma_{2,a}, \tilde{\gamma}_{1,a}$, ${\sigma}_a, \tilde{\sigma}_a$ 
and $z_{i}$, respectively.
We can eliminate    $e_1(\gamma_2)$ from ${p}_2=0$ by using ${p}_{1}=0$.
Then we obtain 
\begin{align}
\begin{split}
& (1-{\sf q}) [ e_2 ({\gamma}_2)+e_{2} ({\gamma}_1)- (e_1 ({\gamma}_2))^2 +e_1 (z) e_1 (\gamma_2)-e_2 (z)] \\
&~~~~~~~~~~~~~~~~~~= -(\lambda_1-\lambda_2-2){\sf q} h e_1 (\gamma_2)+(\lambda_1-\lambda_2-1){\sf q} h e_1 (z) \\
&~~~~~~~~~~~~~~~~~~~~~~~~~+\frac{{\sf q}h^2}{2} (\lambda^2_1+\lambda^2_2-2 \lambda_1\lambda_2-\lambda_1-\lambda_2)+ \frac{(\lambda_1-\lambda_2-1)(\lambda_1-\lambda_2) {\sf q} {}^2 h^2}{1-{\sf q}}  	\,	
\end{split}
\end{align}
where $\lambda_1=r$ and $\lambda_2=n-r$.
From \eqref{eq:generators}, \eqref{eq:parameters1}, and \eqref{eq:paradual}, this provides
\begin{align}
\begin{split}
 &(1-{\sf q}) \langle  e_2 (\sigma)  \rangle^{N_c=r, \, N_f=n}_{\rm A \mathchar `-twist} \\
&= \Big\langle  ({\sf q}-1) [ e_2 (\tilde{\sigma}+h)- (e_1 (\tilde{\sigma}+h))^2 +e_1 (z) e_1 (\tilde{\sigma}+h)-e_2 (z)] 
-(\lambda_1-\lambda_2-2){\sf q} h e_1 (\tilde{\sigma}+h) \\
&+(\lambda_1-\lambda_2-1){\sf q} h e_1 (z) +\frac{{\sf q}h^2}{2} (\lambda^2_1+\lambda^2_2-2 \lambda_1\lambda_2-\lambda_1-\lambda_2) 
+ \frac{(\lambda_1-\lambda_2-1)(\lambda_1-\lambda_2) {\sf q} {}^2 h^2}{1-{\sf q}}  \Big\rangle^{N_c=n-r, \, N_f=n}_{\rm A \mathchar `-twist}
\end{split}
\label{as-sigma}
\end{align}
for correlators of a dual pair of the GLSM.
As before, we can evaluate the LHS for $|q| < 1$.
$\langle e_l (\tilde{\sigma}+h) \rangle^{N_c=n-r, \, N_f=n}_{\rm A \mathchar `-twist}$ on the RHS doesn't have $\tilde{q}$ corrections and \eqref{as-sigma} is written as
\begin{align}
\begin{split}
 \langle  e_2 (\sigma)  \rangle^{N_c=r, \, N_f=n}_{\rm A \mathchar `-twist}
&=\sum_{I_{n-r} \subset I} \prod_{a \in I_{n-r}} \prod_{b \in I_{r}}  \frac{  (-1)^{ r(n-r)} (1-{\sf q})^{-1}}{(z_a-z_b) (z_a-z_b+h) } \\
&~~ \times   \Big(  ({\sf q}-1) [ e_2 (\mathbf{z})- (e_1 (\mathbf{z}))^2 +e_1 (z) e_1 (\mathbf{z})-e_2 (z)] \\
&~~~~~~~~-(\lambda_1-\lambda_2-2){\sf q} h e_1 (\mathbf{z}) +(\lambda_1-\lambda_2-1){\sf q} h e_1 (z)\\
&~~~~~~~~~~+\frac{{\sf q}h^2}{2} (\lambda^2_1+\lambda^2_2-2\lambda_1\lambda_2-\lambda_1-\lambda_2) 
+ \frac{(\lambda_1-\lambda_2-1)(\lambda_1-\lambda_2) {\sf q} {}^2 h^2}{1-{\sf q}}  \Big) \, .
\end{split}
\end{align}
Here $e_l (\mathbf{z})$ means $e_l (z_{i_1}, \cdots, z_{i_{n-r}})$ with $I_{n-r}=\{i_1, \cdots,i_{n-r} \}$.
For example, when $(\lambda_1,\lambda_2):= (r, n-r)=(2,1)$,
\begin{align}
\begin{split}
\langle  e_2 (\sigma)  \rangle^{N_c=2, \, N_f=3}_{\rm A \mathchar `-twist}
&=\sum_{I_{1} \subset I} \prod_{a \in I_1 } \prod_{b \in I_{r}}  \frac{1}{(z_a-z_b) (z_a-z_b+h) } \\
& \times   \Big(    (e_1 (\mathbf{z}))^2 - e_1 (z) e_1 (\mathbf{z}) +e_2 (z)  + \frac{{\sf q} h }{1-{\sf q}} e_1 (\mathbf{z})  -\frac{{\sf q} h^2 }{1-{\sf q}}  \Big) \, , 
\end{split}
\end{align}
and we checked that this agrees with the GLSM calculation.
In a similar manner, we can eliminate  $e_{l} (\gamma_{1}), (l=1, \cdots, k-1)$ from $p_{k}=0$ by using $p_l=0,  (l=1, \cdots, k-1)$ and recursively derive relations between 
$e_{k} (\sigma)$ and $e_{l} (\tilde{\sigma}), (l=1, \cdots, k)$.


\section{Wilson loops in the 3d $\mathcal{N}=2^*$ theory and the Bethe subalgebra of the XXZ$_{1/2}$ model}
In the previous section, we saw that the twisted chiral ring relation of the GLSM agrees with the equivariant quantum cohomology ring of the cotangent bundle of the Grassmannian, which corresponds to the Bethe subalgebra of the XXX$_{1/2}$ spin chain model. 
Therefore we can do similar calculations and checks for the $S^1$ uplift of the twisted chiral ring relation of the 3d $\mathcal{N}=2^*$ theory and the Bethe subalgebra of the XXZ$_{1/2}$ spin chain model \cite{RTV}. 
The $S^1$ uplift of the twisted chiral ring in the 3d $\mathcal{N}=2^*$ theory on $S^1 \times S^2$ is generated by Wilson loops wrapped on $S^1$.

In the 3d $\mathcal{N}=2^*$ theory, which is obtained by the adjoint mass deformation of the 3d $\mathcal{N}=4$ theory, there is a superpotential
\begin{align}
W_{\widetilde{Q} \Phi Q} = \sum_{a,b=1}^{N_c} \sum_{i=1}^{N_f} \widetilde{Q}^a_{i} \Phi^{b}_a Q^{i}_b	\,	.
\end{align}
This breaks $SU(N_f)_Q \times SU(N_f)_{\widetilde{Q}} \times U(1)_D$ to $SU(N_f) \times U(1)_D$. 

\begin{table}[h]
\begin{center}
\begin{tabular}{c | c c c c c }
			&	$U(N_c)$			&	$SU(N_f)$			&	$U(1)_D$	&	$U(1)_T$	&	$U(1)_R$	\\
			\hline
$Q$			&	$N_c$			&	$\overline{N}_f$								&	$-1/2$	&	0			&	$0$		\\
$\widetilde{Q}$	&	$\overline{N}_c$				&	$N_f$				&	$-1/2$	&	0			&	$0$		\\
$\Phi$		&	\text{adj}					&	$\mathbf{1}$					&	$1$		&	0			&	$2$
\end{tabular}
 \caption{Matter contents of the 3d $\mathcal{N}=2^*$ theory}
\label{table:3dv2}
\end{center}
\end{table}

Here, we turn off all the background magnetic fluxes for flavor symmetries.
Then the expectation value of supersymmetric Wilson loops in the representation $\mathcal{R}$ is given by
\begin{align}
\begin{split}
\langle W_{\mathcal{R}} \rangle^{N_c, N_f}_{\text{top. twisted}}  =& \frac{1}{N_c !} \sum_{\vec{m} \in \mathbb{Z}^{N_c}} \oint \prod_{a=1}^{N_c} \frac{d x_a}{2\pi i x_a} 
\mathrm{Tr}_{\mathcal{R}} (x)
 \prod_{a \neq b} \left( 1-  \frac{x_a}{x_b}\right) \prod_{a, b} \left( \frac{x_a^{1/2} x_b^{-1/2} z^{1/2} }{1-x_a x_b^{-1} z} \right)^{m_a-m_b-1} \\
& \qquad \qquad  \times \prod_{a=1}^{N_c} \zeta^{m_a} \prod_{i=1}^{N_f}  
\left( \frac{x_a^{1/2} y_i^{-1/2} z^{-1/4}}{1 -x_a y^{-1}_i z^{-1/2}} \right)^{m_a  +1} 
\left( \frac{x_a^{-1/2} {y}_i^{1/2} z^{-1/4}}{1 -x_a^{-1} {y}_i z^{-1/2}} \right)^{-m_a +1} 	\, ,	\label{WLoop}
\end{split}
\end{align} 
where we absorbed $(-1)^{N_c-1}$ into the definition of the fugacity $\zeta$ for $U(1)_T$. 
Here the fugacity for $SU(N_f)$ flavor symmetry is denoted by $y_i$ and the one for $U(1)_D$ flavor symmetry by $z$.   
When the representation $\mathcal{R}$ is the $l$-th anti-symmetric representation $A_{l}$, $\mathrm{Tr}_{A_l} (x)$ is given by the $l$-th elementary symmetric polynomial of $(x)=\mathrm{diag} (x_1, \cdots,x_{N_c})$
\begin{align}
\mathrm{Tr}_{A_k} (x)=\sum_{1 \le a_1 < \cdots < a_l \le N_c} x_{a_1} \cdots x_{a_l} \, .
\end{align}
Note that any product of supersymmetric Wilson loops is a symmetric function of $(x)$, which is also expressed in terms of the elementary symmetric polynomials.


\subsection{The Bethe subalgebra of the XXZ$_{1/2}$ spin chain model}
It was shown in \cite{RTV} that the Bethe subalgebra of the XXZ$_{1/2}$ model is given by the algebra $\mathcal{K}^{{\sf q}}$,
\begin{align}
\mathcal{K}^{{\sf q}}
\, = \, \mathbb{C}[{\sf z}^{\pm}, \Gamma^{\pm}, {\sf h}^{\pm}]^{S_{\lambda_1} \times S_{\lambda_2} } \otimes \mathbb{C}[[{\sf q}]] \, / \, \mathcal{I}_{{\sf q}}
 \label{eq:QK}
\end{align}
where $\mathbb{C}[\mathbf{z}^{\pm}, \Gamma^{\pm}, {\sf h}^{\pm}]$ is a Laurent polynomial ring of ${\sf z}^{\pm}:=\{ {\sf z}^{\pm}_1, \cdots, {\sf z}^{\pm}_n \}$ and $\Gamma^{\pm}=(\Gamma^{\pm}_1,\Gamma^{\pm}_2)$ with  $\Gamma^{\pm}_1:=\{ \gamma_{1,1}, \cdots, \gamma_{1, \lambda_1} \}$ and $\Gamma^{\pm}_2:=\{ \gamma_{2,1}, \cdots, \gamma_{2,\lambda_2} \}$. Here we take $(\lambda_1, \lambda_2)=(r, n-r)$.
 The $S_{\lambda_i}$ in the exponent denotes the symmetrization of variables $\Gamma^{\pm}_i$. 
 Also, the ideal $\mathcal{I}_{{\sf q}}$ is generated by $n$ coefficients of the following polynomial $P$ of $u^{-1}$
\begin{align}
\begin{split}
&P(\Gamma, {\sf z},{\sf h}, {\sf q}):=(1-{\sf q}) \prod_{i=1}^n (1- u^{-1} {\sf z}_{i})  \\
&~~~~~~~~~~~~~ \qquad - \prod_{i=1}^{2} \prod_{a=1}^{\lambda_i} (1-u^{-1} \gamma_{i,a})
+{\sf q} \prod_{a=1}^{\lambda_1} (1-u^{-1} {\sf h}^{-1} \gamma_{1,a}) \prod_{b=1}^{\lambda_2} (1-u^{-1} {\sf h} \gamma_{2,b}) \, .
 \label{Ideal-XXZ}
\end{split}
\end{align}
So far we do not know the rigorous geometric interpretation of \eqref{eq:QK}, but it was conjectured in \cite{RTV, Bullimore:2014awa} that the algebra $\mathcal{K}^{{\sf q}}$ is  isomorphic to the equivariant quantum $K$-theory ring, $QK^*_{GL_n (\mathbb{C}^*) \times \mathbb{C}^*} (T^* \mathrm{Gr}(r,n), \mathbb{C})$, of the cotangent bundle of the Grassmannian.


\subsubsection*{Bethe ansatz equation and match of parameters}
We consider the identification between generators of $\mathcal{K}^{\sf q}$ and variables in the topologically twisted 3d $\mathcal{N}=2^*$ supersymmetric theory by deriving the Bethe ansatz equation from \eqref{Ideal-XXZ}.

By substituting $u=\gamma_{1,a}$ and $ \gamma_{1,a} {\sf h}^{-1}$ into $P(\Gamma,\mathbf{z},h,q)=0$, we obtain, respectively,
\begin{align}
&(1-{\sf q}) \prod_{i=1}^n (1- \gamma^{-1}_{1,a} {\sf z}_{i})  
=-{\sf q} \prod_{b=1}^{\lambda_1} (1- {\sf h}^{-1} \gamma_{1,a}^{-1}  \gamma_{1,b}) \prod_{b=1}^{\lambda_2} (1- {\sf h} \gamma_{1,a}^{-1}  \gamma_{2,b}) 	\,	,
\label{eq:QK1}
\\
&(1-{\sf q}) \prod_{i=1}^n (1- {\sf h} \gamma_{1,a}^{-1}  {\sf z}_{i}) = \prod_{i=1}^2 \prod_{b=1}^{\lambda_i} (1-{\sf h}  \gamma_{1,a}^{-1} \gamma_{i,b})	\,	.
\label{eq:QK2}
\end{align}
Dividing \eqref{eq:QK1} by \eqref{eq:QK2}, we get the Bethe ansatz equations, 
\begin{align}
\begin{split}
& \prod_{i=1}^n \frac{\gamma_{1,a}-  {\sf z}_{i}}{\gamma_{1,a}- {\sf h}   {\sf z}_{i}} = 
 {\sf h}^{-1} {\sf q} \prod_{b=1 \atop b \neq a}^{\lambda_1}  \frac{\gamma_{1,a}-  {\sf h}^{-1}  \gamma_{1,b}}{\gamma_{1,a}- {\sf h}  \gamma_{1,b}}	\,	,
\label{QKBethe1}
\end{split}
\end{align}
which is the SUSY vacua condition $\zeta e^{i B_a}=1$ of the 3d $\mathcal{N}=2^*$ theory with the following identifications
\begin{align}
 r=N_c, ~~ n=N_f, ~~ \gamma_{1,a}=x_a, ~~ {\sf z}_i=y_i z^{\frac{1}{2}}, ~~ {\sf h}=z^{-1}, ~~ {\sf q}= z^{\frac{N_f}{2}-N_c} \zeta \, .
\label{eq:3dparameters1}
\end{align}


\subsection{Properties of Wilson loop expectation values in the topologically twisted 3d $\mathcal{N}=2^*$ theory }
Here we perform explicit calculations of the Wilson loop expectation value \eqref{WLoop} for several examples and see that they indeed satisfy the Bethe subalgebra $\mathcal{K}^\mathsf{q}$ of XXZ$_{1/2}$ model \eqref{eq:QK}.


\subsubsection*{Abelian cases}
From \eqref{QKBethe1}, which is equivalent to $\zeta e^{i B_a}=1$, we expect that 
the supersymmetric Wilson loop $W=x$ for $U(1)$ gauge theories satisfy
\begin{align}
 \langle \prod_{i=1}^{n} (W-  {\sf z}_{i}) - {\sf q} {\sf h}^{-1} (W-  {\sf h} {\sf z}_{i})\rangle^{N_c=1,N_f=n} =0 
\label{eq:algWilsonAb}
\end{align}
with the parameter identification \eqref{eq:3dparameters1}.

Also by using \eqref{QKBethe1}, the higher order correlation functions $\langle W^{l}  \rangle^{N_c=1,N_f}$ for $l \ge n$ are expressed in terms of $W^k, (k=0,1 \cdots, n-1)$ as 
\begin{align}
\langle W^{l}  \rangle^{N_c=1,N_f=n}  = \sum_{k=0}^{l-1} A^{(k)}_{l}({\sf z},{\sf h}, {\sf q}) \langle W^{l}  \rangle^{N_c=1,N_f=n}	\,	.
\end{align}
In the 2d $\mathcal{N}=(2,2)^*$ theory with $N_f=n$ flavors, we found $\langle \sigma^l  \rangle^{N_c=1,N_f=n}_{\text{A-twist}}$ with $l \leq n-1$ do not have $q$ corrections.
There is a similar property in the 3d $\mathcal{N}=2^*$ theory. 
For $0 \le l \le n-1$, $\langle W^{l} \rangle^{N_c=1,N_f=n}$ does not have $\zeta$ corrections and is given by the zero magnetic charge sector    
\begin{align}
&\langle W^{l} \rangle^{N_c=1,N_f=n}
=\sum_{i=1}^{n} \prod_{j=1 \atop j \neq i}^{n}
\frac{(-1)^{n-1} {\sf z}^l_i }{ \left(( {\sf z}_i {\sf z}^{-1}_j)^{\frac{1}{2}}-( {\sf z}_i {\sf z}^{-1}_j)^{-\frac{1}{2}} \right) 
 \left( ({\sf h}^{-1} {\sf z}_i {\sf z}^{-1}_j)^{\frac{1}{2}}- ({\sf h}^{-1} {\sf z}_i {\sf z}^{-1}_j)^{-\frac{1}{2}} \right) } \, . 
\label{eq:3deqint}
\end{align}
We have checked \eqref{eq:algWilsonAb} and \eqref{eq:3deqint} for $N_f=2,3,4$ in several orders of $\zeta$.


\subsubsection*{Non-Abelian cases}
In two dimensions, we observed that the partition function $\langle 1 \rangle_{\text{A-twist}}$ does not receive any $q$ corrections and is given by residues at the zero magnetic charge sector. 
Similarly we observed that the partition function (index) of the topologically twisted 3d $\mathcal{N}=2^*$ theory on $S^1 \times S^2$ does not receive $\zeta$ corrections neither and is given by
\begin{align}
&\langle 1 \rangle^{N_c=r,N_f=n}
=\sum_{I_r \subset I} \prod_{i \in I_r} \prod_{j \in I_{n-r}} 
\frac{(-1)^{(n-r)r}  }{ \left(( {\sf z}_i {\sf z}^{-1}_j)^{\frac{1}{2}}-( {\sf z}_i {\sf z}^{-1}_j)^{-\frac{1}{2}} \right) 
 \left( ({\sf h}^{-1} {\sf z}_i {\sf z}^{-1}_j)^{\frac{1}{2}}- ({\sf h}^{-1} {\sf z}_i {\sf z}^{-1}_j)^{-\frac{1}{2}} \right) }	\,	.
\label{eq:indexres}
\end{align}
In two dimensions, $\langle 1 \rangle^{N_c=r,N_f=n}_{\text{A-twist}}$ has a geometrical interpretation as the equivariant integration of $[1] \in H^* (T^*\mathrm{Gr} (r,n);\mathbb{C})$. 
The index \eqref{eq:indexres} also has a geometrical interpretation.
If we identify 3d parameters ${\sf z}_i$ and ${\sf h}$ as ${\sf z}_i=e^{z_i}$ and ${\sf h}=e^{-h}$, respectively, \eqref{eq:indexres} is the sinh uplift of the equivariant integration, which can be interpreted as the equivariant Dirac index.

For $2 \leq n-2$ $(r=2)$, we also observe that the expectation values of $x^{\pm1}_1+x^{\pm1}_2$, $(x_1 x^{-1}_2)^{\pm1}+(x^{-1}_1 x_2)^{\pm1}$, and $(x_1 x_2)^{\pm 1}$ do not have $\zeta$ corrections.
For example, the expectation value of $W^{\pm1}_{F}=x^{\pm1}_1+x^{\pm1}_2$ is
\begin{align}
&\langle W^{\pm1}_{F}  \rangle^{N_c=2,N_f=n}
=\sum_{I_2 \subset I} \prod_{i \in I_2} \prod_{j \in I_{n-2}}  
\frac{ (\sum_{a \in I_2} {\sf z}^{\pm}_a )  }{ \left(( {\sf z}_i {\sf z}^{-1}_j)^{\frac{1}{2}}-( {\sf z}_i {\sf z}^{-1}_j)^{-\frac{1}{2}} \right) 
 \left( ({\sf h}^{-1} {\sf z}_i {\sf z}^{-1}_j)^{\frac{1}{2}}- ({\sf h}^{-1} {\sf z}_i {\sf z}^{-1}_j)^{-\frac{1}{2}} \right) } \, .
\end{align}
However, the properties of correlation functions $(x_1+x_2)^2$, $(x_1+x_2) (x_1 x_2)$, and $(x_1 x_2)^2$ are different from the 2d case.
In the 2d $\mathcal{N}=(2,2)^*$  theory with $N_c \leq N_f-N_c$, we expected that correlation functions of symmetric polynomials of $\sigma$, $\langle \prod_{a=1}^{N_c} e^{l_a}_a(\sigma) \rangle$ with $\sum_{a=1}^{N_c} l_a \le N_c$, don't have $q$ dependence. 
This may be because the degree of the polynomial cannot be reduced to a lower degree in the polynomial ring by the ideal.
For example, $\langle (\sigma_1+\sigma_2)^l (\sigma_1 \sigma_2)^{k} \rangle^{N_c=2, \, N_f=4}$ with $k+l=2$ agrees with the residues at zero magnetic flux sector and do not have $q$ dependence.
On the other hand, if we eliminate $\gamma_{2,1}+\gamma_{2,2}$ and $\gamma_{2,1} \gamma_{2,2}$ from the ideal of $\mathcal{K}^{{\sf q}}$, we obtain
\begin{align}
&\frac{ (1-{\sf h}^2 {\sf q} ) e_4 ({\sf z}) }{e_2 (\gamma_1)}+e_2 (\gamma_1)
    (1-\frac{\sf q}{{\sf h}^2} )+\frac{ ({\sf q}-1)  [ e^2_1 (\gamma_1) ({\sf q}-{\sf h})-{\sf h} ({\sf q}-1) e_1 (\gamma_1) e_1({\sf z})  ] }{ {\sf h} ({\sf h} {\sf q}-1)}+({\sf q}-1) e_{2} ({\sf z})  =0 	\,	,
\label{eq:recursion1} \\
&\frac{( {\sf h} {\sf q}-1) e_4 ({\sf z}) e_1 (\gamma_1)}{e_2 (\gamma_1)} 
  - ( {\sf h} - {\sf q} ) e_2 (\gamma_1) \frac{( {\sf h} - {\sf q} ) e_1 (\gamma_1) +{\sf h} ({\sf q}-1) e_1 ({\sf z}) }{ {\sf h}^2 ({\sf h} {\sf q}-1)}+(1-{\sf q} e_3 ({\sf z}))=0 \, .
\label{eq:recursion2}
\end{align}
Then we find that the degree of $\langle (x_1+x_2)^l (x_1 x_2)^{k} \rangle^{N_c=2,N_f=4}$ with $k+l=2$ is reduced by the above equations and that $(x_1+x_2)^l (x_1 x_2)^{k} $ has $\zeta$ dependence.
We have checked \eqref{eq:recursion1} and \eqref{eq:recursion2} hold for several orders of $\zeta$ in terms of expectation values of Wilson loops.


\subsection{Wilson loops and the Seiberg-like duality in the 3d $\mathcal{N}=2^*$ theory}
We proceed similarly for $N_c > N_f - N_c$ as we did for the 2d $\mathcal{N}=(2,2)^*$ case.
We consider the polynomial $P(\tilde{\Gamma},\tilde{\mathbf{z}}, \tilde{{\sf h}}, \tilde{{\sf q}})=0$ from the ideal for the case $(\tilde{\lambda}_1, \tilde{\lambda}_2)=(n-r,n)$,
\begin{align}
\begin{split}
(1-\tilde{{\sf q}}) \prod_{i=1}^n (1- u^{-1} \tilde{{\sf z}}_{i}) = \prod_{i=1}^{2} \prod_{a=1}^{\tilde{\lambda}_i} (1-u^{-1} \tilde{\gamma}_{i,a})
-
{\sf q} \prod_{a=1}^{\tilde{\lambda}_1} (1-u^{-1} \tilde{{\sf h}}^{-1} \tilde{\gamma}_{1,a}) \prod_{b=1}^{\tilde{\lambda}_2} (1-u^{-1} \tilde{{\sf h}} \tilde{\gamma}_{2,b})	\,	.
\label{dualidealXXZ2}
\end{split}
\end{align}
With the identifications
\begin{align}
 \lambda_1=\tilde{\lambda}_2, ~~ \lambda_2=\tilde{\lambda}_1, ~~ \gamma_{1,a}=\tilde{\gamma}_{2,a} \tilde{{\sf h}}, 
~~ \gamma_{2,a}=\tilde{\gamma}_{1,a} \tilde{{\sf h}}^{-1}, ~~ {\sf z}_i=\tilde{{\sf z}}_{i},~~ {\sf h}=\tilde{{\sf h}}, ~~ {\sf q}= \tilde{{\sf q}}^{-1} \, 	, 
\label{eq:3dparameters3}
\end{align}
\eqref{dualidealXXZ2} is identical to $P(\Gamma,\mathbf{z},{\sf h}, {\sf q})$ for $(\lambda_1, \lambda_2)=(r,n-r)$. 
Thus, with \eqref{eq:3dparameters3} the Bethe subalgebra for $(\lambda_1, \lambda_2)=(r,n-r)$ and the one for $(\tilde{\lambda}_1, \tilde{\lambda}_2)=(n-r, n)$ are isomorphic.
By substituting $u= \tilde{\gamma}_{1,a}, \tilde{\gamma}_{1,a} \tilde{{\sf h}}^{-1}$ into \eqref{dualidealXXZ2}, we obtain
\begin{align}
\begin{split}
& \prod_{i=1}^n \frac{\tilde{\gamma}_{1,a}-  \tilde{{\sf z}}_{i}}{\tilde{\gamma}_{1,a}- \tilde{{\sf h}}   \tilde{{\sf z}}_{i}} \, = \,  
 \tilde{{\sf h}}^{-1} \tilde{{\sf q}} \prod_{b=1 \atop b \neq a}^{\tilde{\lambda}_1}  \frac{\tilde{\gamma}_{1,a}-  \tilde{{\sf h}}^{-1}  
\tilde{\gamma}_{1,b}}{\tilde{\gamma}_{1,a}- \tilde{{\sf h}}  \tilde{\gamma}_{1,b}} \, ,
\label{QKdualBethe1}
\end{split}
\end{align}
which is again the same with the SUSY vacua condition $\tilde{\zeta} e^{i B_a}=1$ of the $U(n-r)$ gauge theory with the identifications
\begin{align}
 n-r=N_c, ~~ n=N_f, ~~ \tilde{\gamma}_{1,a}=\tilde{x}_a, ~~ \tilde{{\sf z}}_i=\tilde{y}_i \tilde{z}^{\frac{1}{2}}, ~~ 
\tilde{{\sf h}}=\tilde{z}^{-1}, ~~ \tilde{{\sf q}}= \tilde{z}^{\frac{N_f}{2}-N_c} \tilde{\zeta} \, . 
\label{eq:3dparameters4}
\end{align}
From \eqref{eq:3dparameters1}, \eqref{eq:3dparameters3} and \eqref{eq:3dparameters4}, we have maps between parameters in $U(r)$ and $U(n-r)$ 3d $\mathcal{N}=2^*$ gauge theories,
\begin{align}
y_i  =\tilde{y}_i , ~~ \ \ z =\tilde{z}, ~~ \ \  \zeta= {\tilde{\zeta}}^{-1} 	\,	. 
\end{align}
So from now on, we don't distinguish $y_i$, $z$, and ${\sf z}_i$ from $\tilde{y}_i$, $\tilde{z}$, and $\tilde{{\sf z}}_i$, respectively.  	\\

From \eqref{eq:indexres}, we have 
\begin{align}
\langle 1  \rangle^{N_c, \, N_f}=\langle 1  \rangle^{N_f-N_c, \, N_f}
\end{align}
at the level of the partition function (or index). We have checked this for several $N_f$ and $N_c$. 	\\

For the fundamental representation, we consider the coefficient of $u^{-n+1}$ of  $P=0$,
\begin{align}
(1- {\sf h} {\sf q}) \sum_{a=1}^{\lambda_2} \gamma_{2,a}=   -(1- {\sf h}^{-1} {\sf q}) (\sum_{a=1}^{\lambda_1} \gamma_{1,a}) 
-(1-{\sf q}) \left(\sum_{i=1}^n {\sf z}_{i} \right)	\, ,
\label{eq:coeffQK1}
\end{align}
which becomes
\begin{align}
(1-  z^{\frac{n}{2}-r-1} \zeta) z \left( \sum_{a=1}^{n-r} \tilde{x}_{a} \right) 
=   -(1- z^{\frac{n}{2}-r+1} \zeta) \left(\sum_{a=1}^{r} x_{a} \right) -(1-z^{\frac{n}{2}-r} \zeta)\left(\sum_{i=1}^n {\sf z}_{i} \right)
\end{align}
by using relations between two sets of parameters, \eqref{eq:3dparameters1}, \eqref{eq:3dparameters3} and \eqref{eq:3dparameters4}.
Therefore, this indicates that the Wilson loop in the fundamental representation $W_{F}=\sum_{a=1}^{N_c} x_a$ in the $U(N_c)$ gauge theory with $N_c > N_f - N_c$ is provided by
\begin{align}
\begin{split}
&(1- z^{\frac{n}{2}-r+1} \zeta) \langle W_F \rangle^{N_c=r, N_f=n}
+(1-z^{\frac{n}{2}-r} \zeta) \left(\sum_{i=1}^n {\sf z}_{i}  \right) \langle 1 \rangle^{N_c=r, N_f=n}	\\
&	\qquad	\qquad	\qquad	\qquad	\qquad	\qquad	\qquad	\qquad	\qquad		
= - (1- z^{\frac{n}{2}-r-1} \zeta ) z \langle \widetilde{W}_F  \rangle^{N_c=n-r, N_f=n} 
\end{split}
\end{align}
where $\widetilde{W}_F  = \sum_{a=1}^{n-r} \tilde{x}_{a}$ is the Wilson loop in the fundamental representation in the $U(N_f-N_c)$ gauge theory.

When calculating the index, the evaluation of the LHS in the region $\zeta <1$ (resp. $\zeta >1$) means that the RHS is evaluated in the region $\tilde{\zeta}=\zeta^{-1} >1$ (resp. $\tilde{\zeta}=\zeta^{-1} <1$) where the negative (resp. positive) magnetic fluxes contribute to the Jeffrey-Kirwan residue operations.
We evaluated the LHS and the RHS separately and have agreement for several $r$ and $n$.	\\

Next we consider the second antisymmetric representation. 
We eliminate $e_{1}(\gamma_1)$ from the coefficient of $u^{-2}$ in $P=0$.
Then we obtain the relation
\begin{align}
\begin{split}
&(1-  {\sf h}^{-2} {\sf q}) e_2( \gamma_1)+(1-  {\sf h}^{2} {\sf q}) e_2( \gamma_2)
 =(1-   {\sf q}) \Big[ \frac{(1-  {\sf h} {\sf q})}{(1-  {\sf h}^{-1} {\sf q})} (e_1( \gamma_2))^2 - \frac{(1-  {\sf q})}{(1-  {\sf h}^{-1} {\sf q})}  e_1(\sf{z}) e_1( \gamma_2) + e_2(\sf{z}) \Big] \, .
\end{split}
\end{align}
This can be written in terms of $x_a$ and $\tilde{x}_a$ as
\begin{align}
\begin{split}
&(1-  {\sf h}^{-2} {\sf q}) e_2( x )=
-(1-  {\sf h}^{2} {\sf q}) e_2( \tilde{x} {\sf h}^{-1} ) \\
&~~~~~~~~~~~~~~~~+(1-   {\sf q}) \Big[ \frac{(1-  {\sf h} {\sf q})}{(1-  {\sf h}^{-1} {\sf q})} (e_1(\tilde{x} {\sf h}^{-1} ))^2 \textcolor{blue}{-} \frac{(1-  {\sf q})}{(1-  {\sf h}^{-1} {\sf q})} e_1({\sf z}) e_1( \tilde{x} {\sf h}^{-1}) +e_2({\sf z}) \Big] \, .
\end{split}
\end{align}
Therefore, this suggests that the expectation value of the second antisymmetric representation $W_{A_2}=\sum_{a < b} x_a x_b$ is given by
\begin{align}
\begin{split}
&(1-  {\sf h}^{-2} {\sf q}) \langle W_{A_2} \rangle^{N_c=r, N_f=n}  \\
 &=
\Big\langle ({\sf q}-  {\sf h}^{2} )   \widetilde{W}_{A_2}   
+(1-   {\sf q}) \Big[ \frac{({\sf h}^{-1} -  {\sf q})}{({\sf h}-   {\sf q})} (\widetilde{W}_{F})^2 - \frac{(1-  {\sf q})}{({\sf h}-  {\sf q})} e_1({\sf z}) \widetilde{W}_{F}  +e_2({\sf z}) \Big] 
\Big\rangle^{N_c=n-r, N_f=n}
\end{split}
\end{align}
where $\widetilde{W}_{A_2} = \sum_{a < b} \tilde{x}_a \tilde{x}_b$, $\widetilde{W}_F = \sum_a \tilde{x}_a$, $\mathsf{q} = z^{\frac{N_f}{2}-N_c} \zeta$, and $\mathsf{h}=z^{-1}=\tilde{z}^{-1}$.
We checked this for several $N_c$ and $N_f$.

In a similar way, we can have the Seiberg-like duality for Wilson loops in other representations from the ideal with identification of parameters \eqref{eq:3dparameters1}, \eqref{eq:3dparameters3}, and \eqref{eq:3dparameters4}.
Also, as done in the 2d case, we can have the $S^1$-uplift of twisted chiral rings by eliminating symmetric polynomials of $\gamma_{a,2}$ in \eqref{Ideal-XXZ} with the identification of parameters \eqref{eq:3dparameters}.


\section{Conclusion and future directions}
\label{sec:concl}
In this paper, we discussed the relation between the partition function in the A-twisted 2d $\mathcal{N}=(2,2)$ theory (resp. the topologically twisted 3d $\mathcal{N}=2$ gauge theory) and the inverse of the norm of the Bethe eigenstate for the XXX$_{1/2}$ (resp. XXZ$_{1/2}$) spin chain model with a particular choice of $R$-charges and background magnetic fluxes for flavor symmetries in the gauge theory side. 
Coefficients of the expectation value of the Baxter $Q$-operator and the conserved charges were understood in terms of correlation functions in gauge theories.

We also studied the relation between correlation functions in the A-twisted 2d $\mathcal{N}=(2,2)^*$ $U(N_c)$ gauge theories and the equivariant integration of equivariant quantum cohomology classes for the cotangent bundle of the Grassmannian. 
We calculated each of them for several examples, checked that they agree, and expect that the relation holds for general cases.
For the case $N_c > N_f- N_c$, we used the isomorphism of Grassmannians to calculate the equivariant integration where such isomorphism corresponds to the Seiberg-like duality in the GLSM side.

As the twisted chiral ring of the 2d $\mathcal{N}=(2,2)^*$ theory is identified with the Bethe subalgebra of the XXX$_{1/2}$ spin chain model, we were able to make a similar identification for the 3d $\mathcal{N}=2^*$ theory. 
We calculated correlation functions of Wilson loops and checked that they agree with the Bethe subalgebra of the XXZ$_{1/2}$ spin chain model. \\

There are several interesting directions. 
Firstly, it will be interesting to find the analogue of the equivariant integration in the equivariant quantum $K$-theory and match them with the correlation functions of Wilson loops in the topologically twisted 3d $\mathcal{N}=2^*$ theory.

Another interesting direction is to study relations between the Bethe ansatz and the finite-dimensional commutative Frobenius algebra.
In \cite{Korff:2013rsa}, a finite-dimensional commutative Frobenius algebra was constructed in terms of the Bethe ansatz for the $q$-boson model. 
It is known that the finite-dimensional commutative Frobenius algebra is essentially same as the 2d topological quantum field theory (TQFT) and the 2d partition function on genus $g$ Riemann surface $\Sigma_g$ corresponding to $q$-boson can be written as \cite{Okuda-Yoshida}
\begin{align}
Z(\Sigma_g)=\sum_{(\lambda) \in P_{\text{q-boson}}} \langle \Psi (\lambda) | \Psi (\lambda) \rangle^{g-1} .
\end{align}
Here $ | \Psi (\lambda) \rangle$ is the eigenvector of the $q$-boson determined by the Bethe root $(\lambda)$. 
We obtained the same type of formula for the XXX$_{1/2}$ spin chain model where the corresponding TQFT is the topologically twisted 2d $\mathcal{N}=(2,2)$ theory and also for the XXZ$_{1/2}$ model that corresponds to the 3d $\mathcal{N}=2$ theory with the partial topological twist along $S^2$.
By using recent results \cite{Benini:2016hjo, ClossetKim}, the partition function of the 2d $\mathcal{N}=(2,2)$ and the 3d $\mathcal{N}=2$ theories studied in this paper can be generalized to Riemann surfaces of genus $g$ as
\begin{align}
Z(\Sigma_g) &= \sum_{(\lambda) \in P_{\text{XXX}}} \langle \Psi (\lambda) | \Psi (\lambda) \rangle^{g-1} \, , \\
Z(S^1 \times \Sigma_g) &= \sum_{(\lambda) \in P_{\text{XXZ}}} \langle \Psi (\lambda) | \Psi (\lambda) \rangle^{g-1} \, .
\end{align}
These formulas are similar to the $q$-boson case and imply that there exist finite-dimensional commutative Frobenius algebras associated with the Bethe ansatz for the XXX$_{1/2}$ and also the XXZ$_{1/2}$ spin chain models. 
It would be interesting to construct the Frobenius algebras in terms of the XXX$_{1/2}$ and the XXZ$_{1/2}$ spin chain models.

\acknowledgments{We would like to thank Hiroaki Kanno, Bumsig Kim, Heeyeon Kim, Jeongseok Oh, Satoshi Okuda and Kazushi Ueda for interesting discussion. H.-J.C appreciates the IBS Center for Geometry and Physics, the Department of Mathematics at Postech, and the Institut Henri Poincar\'e for hospitality where part of this work was performed.}


\appendix

\section{Reducing the polynomial ring by the ideal in $T^{*}\text{Gr}(2,4)$}
\label{appA}
In this appendix, we provide some detail calculation to get \eqref{eq:Gr24red}. 
From the ideal \eqref{eq:ideal},
\begin{align}
\begin{split}
 \left(u-\gamma _{1,1}\right) \left(u-\gamma _{1,2}\right) \left(u-\gamma _{2,1}\right) \left(u-\gamma _{2,2}\right)-{\sf q} \left(\gamma _{1,1}+h-u\right) \left(\gamma _{1,2}+h-u\right) \left(-\gamma _{2,1}+h+u\right) \left(-\gamma _{2,2}+h+u\right) \\
  - \left(1-{\sf q}\right) \left(u-z_1\right) \left(u-z_2\right) \left(u-z_3\right) \left(u-z_4\right)	\,	,
\end{split}
\end{align}
we obtain four relations $p_l=0, \, l=1,2,3,4$,
\normalsize{
\begin{align}
p_1 \,  
=& \, \left({\sf q}-1\right) \left(e_1(\gamma_1)+e_1(\gamma_2)-e_1( z )\right)	\,	, 
\label{eq:p1} 	\\
p_2 \,  
=& \, (1-{\sf q}) \left( e_1(\gamma _{1}) e_1(\gamma _{2}) + e_2(\gamma _{1})  + e_2(\gamma _{2}) \right) 
 + {\sf q} h \left( e_1(\gamma _{1})  - e_1(\gamma _{2})   + 2 h \right) +\left({\sf q}-1\right) e_2(z)	\,	,     
\label{eq:p2} 	\\
p_3 \,  
=& \, ({\sf q} - 1)\left(e_1(\gamma _{1}) e_2(\gamma _{2}) +e_2(\gamma _{1}) e_1(\gamma _{2})\right) 
-h^2 \left(e_1(\gamma _{1})+e_1(\gamma _{2})\right)-2 h \left(e_2(\gamma _{1} )-e_2(\gamma _{2} )\right) + \left(1-{\sf q}\right)e_3(z)	\,	,	  \\
p_4 \,  
=& \, e_2(\gamma _{1}) e_2(\gamma _{2})-{\sf q} \left(e_2(\gamma_1)+h e_1(\gamma_1)+h^2 \right)\left(e_2(\gamma_2)-h e_1(\gamma_2)+h^2 \right) +e_4(z) \left({\sf q}-1\right)	\,	.
\end{align}
Solving \eqref{eq:p1} and \eqref{eq:p2}, $e_l(\gamma_2)$, $l=1,2$ can be expressed in terms of $e_{1}(\gamma_1)$ and $e_{2}(\gamma_1) $,
\begin{align}
e_1(\gamma_2) \,  =& \,  -e_1(\gamma_1)+e_1( z )	\,	, 
\label{eq:p1s} \\
e_2(\gamma_{2}) \,  =& \, e_1(\gamma _{1})^2 -e_2(\gamma _{1})  -\left(e_1( z )+ \frac{2{\sf q} h}{1-{\sf q}} \right)e_1(\gamma _{1} )  + e_2(z) 
 + \frac{{\sf q} h}{1-{\sf q}} \left(   e_1(z)   - 2 h \right)	\,	.      
\label{eq:p2s} 
\end{align}
Plugging \eqref{eq:p1s} and \eqref{eq:p2s}  into $p_3=0$, $e_1(\gamma_1)^3$ can be written as a function of $e_{k}(\gamma_1) e_{l}(\gamma_1)$  with $k+l \le 2$,
\begin{align}
\begin{split}
e_1(\gamma _{1})^3 \,  =& \, \left( \frac{2h(1+{\sf q})}{1-{\sf q}}+e_1(z) \right)e_1(\gamma_1)^2 +2 e_1(\gamma_1)e_2(\gamma_1) \\
&+\left( \frac{ -2 h^2 {\sf q} ({\sf q}+1)+h \left({\sf q}^2+{\sf q}-2\right)  e_1(z)}{(1-{\sf q})^2} - e_2(z) \right) e_1(\gamma_1)\\
&-\left(e_1(z) +\frac{4 h}{1-{\sf q}}\right) e_2(\gamma_1)
+e_3(z)
+\frac{-4 h^3 {\sf q}+h^2 (3 {\sf q}-1)  e_1(z)+2 h (1-{\sf q}) e_2(z)}{(1-{\sf q})^2}
\label{eq:deg3red1} 
\end{split}
\end{align}
Plugging \eqref{eq:p1s} and \eqref{eq:p2s}  into $p_4=0$, $e_1(\gamma _{1})^2 e_2(\gamma _{1})$ is given by 
\begin{align}
\begin{split}
&\hspace{-10mm}e_1(\gamma _{1})^2 e_2(\gamma _{1})\, =  \, \frac{h {\sf q}}{1- {\sf q}} e_{1}(\gamma_1)^3+
\frac{h {\sf q} (2h(-1+2 {\sf q})+(1-{\sf q})e_1(z))}{(1-{\sf q})^2}e_1(\gamma_1)^2
-\left( e_1(z)+\frac{2h {\sf q} }{1-{\sf q}} \right) e_1(\gamma_1) e_2(\gamma_1)		\\
&\hspace{15mm}-e_2(\gamma_1)^2+\frac{h {\sf q}( (-1+{\sf q}) e_2(z)+h (2-3 {\sf q} e_1(z)+h^2 (-2+6 {\sf q})))}{(1-{\sf q})^2} e_1(\gamma_1)		\\
&\hspace{15mm}+\left( e_2(z) +\frac{2h {\sf q}(e_1(z)-h) }{1-{\sf q}} \right) e_2(\gamma_1) -e_4(z)+\frac{h^2 {\sf q}(h e_1(z)-e_2(z)-h^2)}{1-{\sf q}}
+\frac{h^3 {\sf q}^2(2h - e_1(z))}{(1-{\sf q})^2}	\,	.
\end{split}
\label{eq:deg3red2}
\end{align}
Finally, by plugging \eqref{eq:deg3red1} into \eqref{eq:deg3red2}, we obtain $e_1(\gamma _{1})^2 e_2(\gamma _{1})$ as a function of $e_{l}(\gamma_1) e_{k}(\gamma_1)$  with $k+l \le 2$.

In the same manner, by plugging solutions of $p_1=0, p_4=0 $ into $p_3=0$, $e_1(\gamma _{1})e_2(\gamma _{1})^2$ is expressed as
{\scriptsize
\begin{align}
\begin{split}
\hspace{-20mm}e_1(\gamma _{1})e_2(\gamma _{1})^2 & \,  = 
+\frac{2  h {\sf q}}{1-{\sf q}}e_1^2(\gamma_1) e_2(\gamma_1)
+\frac{ h^2 {\sf q}}{1-{\sf q}}e_1^3(\gamma_1)
+\left(e_1(z)+\frac{2 h}{1-{\sf q}}\right)e_2^2(\gamma_1)
\\
&-\frac{2  h {\sf q} (h {\sf q}+h-{\sf q}e_1(z) +e_1(z))}{(1-{\sf q})^2} e_1(\gamma_1) e_2(\gamma_1)
+\frac{ h^2 {\sf q} (2 h {\sf q}-{\sf q} e_1(z)+e_1(z))}{(1-{\sf q})^2}e_1^2(\gamma_1) 
+\frac{ \left(-4 h^3 {\sf q}+h^2 \left({\sf q}^2+1\right) e_1(z)-(1-{\sf q})^2 e_3(z)\right)}{(1-{\sf q})^2} e_2(\gamma_1)
\\
&+\frac{ \left(-{\sf q}h^4  ({\sf q}+3)+h^3 {\sf q}^2 e_1(z)+h (1-{\sf q}) {\sf q} e_3(z)+(1-{\sf q})^2 e_4(z) \right)}{(1-{\sf q})^2}e_1(\gamma_1)
+\frac{h \left(-2 h^4 {\sf q}+h^3 {\sf q}e_1 (z)-h ({\sf q}-1) {\sf q} e_3 (z)+2 ({\sf q}-1) e_4 (z)\right)}{(1-{\sf q})^2}	\,	.
\label{eq:deg3red3}
\end{split}
\end{align}
}
Also, by using \eqref{eq:deg3red1} and \eqref{eq:deg3red2}, $e_1(\gamma _{1}) e_2(\gamma _{1})^2$ can be reduced to a function of $e_{l}(\gamma_1) e_{k}(\gamma_1)$  with $k+l \le 2$.

Similarly, by plugging the solutions of $p_2=0, p_3=0 $ into $p_4=0$, $e_2(\gamma _{1})^3$ can be written as
{\scriptsize
\begin{align}
\begin{split}
& \hspace{-20mm} e_2(\gamma _{1})^3  \,  =
\frac{ h ({\sf q}+2)}{1-{\sf q}} e_1(\gamma_1) e_2^2 (\gamma_1)
+\frac{ h^2}{1-{\sf q}} e_1^2(\gamma_1) e_2(\gamma_1) 
+ \left(\frac{h^2 \left(2 {\sf q}^2-7 {\sf q}-1\right)}{(1-{\sf q})^2}+e_2(z)\right)e_2^2 (\gamma_1)
+\left(\frac{2 h^3 {\sf q} (2 {\sf q}-5)}{(1-{\sf q})^2}-\frac{2 h {\sf q} e_2(z)}{1-{\sf q}}-e_3(z)\right) e_1(\gamma_1) e_2(\gamma_1) 
\\
&\hspace{-8mm} + \left(\frac{h^4 ({\sf q}-3) {\sf q}}{(1-{\sf q})^2}-\frac{h^2 {\sf q} e_2(z) }{1-{\sf q}}+\frac{h {\sf q} e_3(z)}{1-{\sf q}}+e_4(z) \right) e_1^2(\gamma_1)
+\frac{ \left(h^4 {\sf q} (3 {\sf q}-7)+h^2 \left({\sf q}^2+1\right)  e_2(z)-2 h ({\sf q}-1) {\sf q}e_3(z) -({\sf q}-1)^2 e_4(z) \right)}{(1-{\sf q})^2}e_2(\gamma_1)
\\
&\hspace{-8mm} +\frac{ h \left(h^4 ({\sf q}-4) {\sf q}+h^2 {\sf q}^2 e_2(z)+h (2-3 {\sf q}) {\sf q} e_3(z)+\left({\sf q}^2+{\sf q}-2\right) e_4(z)\right)}{(1-{\sf q})^2}e_1(\gamma_1)
+\frac{h^2 \left(-h^4 {\sf q}+h^2 {\sf q} e_2(z)+h (1-2 {\sf q}) q e_3(z)+(3 {\sf q}-1) e_4(z)\right)}{(1-{\sf q})^2}	\,	.
\label{eq:deg3red4}
\end{split}
\end{align}
}
We obtain $e_2(\gamma_1)^{3}$ as a function of $e_{l}(\gamma_1) e_{k}(\gamma_1)$ with $k+l \le 2$ from \eqref{eq:deg3red1}-\eqref{eq:deg3red4}.


\newpage

\bibliographystyle{JHEP}
\bibliography{toolbox}

\end{document}